\newfont{\vs}{cmssdc10 scaled 1050}

\documentclass[useAMS,usenatbib,usegraphicx]{mn2e}
\usepackage{xspace,amsmath}
\usepackage{color,epsfig,amssymb,lscape}
\usepackage{util}
\usepackage{array,colortbl}
\usepackage[dvips]{graphicx}
\usepackage{times,url,graphicx}
\usepackage{longtable}
\usepackage{txfonts}
\usepackage[figuresright]{rotating}
\usepackage{subfigure}
\usepackage{multirow}
\usepackage{rotating}
\usepackage{afterpage}
\usepackage{lscape}

\title[The extended HeII$\lambda$4686 emission in SBS\,0335-052E]{The extended HeII$\lambda$4686 emission in the extremely
  metal-poor galaxy SBS\,0335-052E seen with MUSE\thanks{Based on observations collected at the ESO}}
\author[C. Kehrig et
al.]{C. Kehrig$^{1}$\thanks{E-mail:kehrig@iaa.es}, J.M. V\'{\i}lchez$^{1}$, M.A. Guerrero$^{1}$, J. Iglesias-P\'aramo$^{1,2}$, L.K. Hunt$^{3}$ 
\newauthor
S. Duarte Puertas$^{1}$, G. Ramos-Larios$^{4}$ \\
$^{1}$ Instituto de Astrof\'{\i}sica de Andaluc\'{\i}a, CSIC, Apartado de correos 3004, 18080 Granada, Spain \\
$^{2}$ Estaci\'on Experimental de Zonas Aridas (CSIC), Ctra. de Sacramento s/n, La Caada, Almer\'{\i}a, Spain \\
$^{3}$ INAF-Osservatorio Astrofisico di Arcetri, Largo E. Fermi, 5, 50125, Firenze, Italy \\
$^{4}$ Instituto de Astronom\'{\i}a y Meteorolog\'{\i}a, Dpto. de F\'{\i}sica, CUCEI, Universidad de Guadalajara \\ Av. Vallarta No. 2602, C.P. 44130, Guadalajara, Jalisco, Mexico}

\begin{document}

\date{Accepted Date. Received Date; in original Date}

\pagerange{\pageref{firstpage}--\pageref{lastpage}} \pubyear{2016}

\maketitle

\label{firstpage}

\begin{abstract}
SBS\,0335-052E, one of the most metal-poor (Z $\sim$ 3-4$\%$
Z$_{\odot}$) HeII-emitter starbursts known in the nearby universe, is
studied using optical VLT/MUSE spectroscopic and Chandra X-ray
observations. We spatially resolved the spectral map of the nebular
HeII$\lambda$4686 emission from which we derived for the first time
the total HeII-ionizing energy budget of SBS\,0335-052E. The nebular
HeII line is indicative of a quite hard ionizing spectrum with photon
energies $>$ 4 Ryd, and is observed to be more common at high-z than
locally.  Our study rules out a significant contribution from X-ray
sources and shocks to the HeII photoionization budget, indicating that
the He$^{+}$ excitation is mainly due to hot stellar continua. We
discovered a new WR knot, but we also discard single WR stars as the
main responsible for the HeII ionization. By comparing observations
with current models, we found that the HeII-ionization budget of
SBS\,0335-052E can only be produced by either single, rotating
metal-free stars or a binary population with Z $\sim$ 10$^{-5}$ and a
'top-heavy' IMF. This discrepancy between the metallicity of such
stars and that of the HII regions in SBS\,0335-052E is similar to
results obtained by \cite{K15} for the very metal-deficient
HeII-emitting galaxy IZw18. These results suggest that the HeII
ionization is still beyond the capabilities of state-of-the-art
models.  Extremely metal-poor, high-ionizing starbursts in the local
universe, like SBS\,0335-052E, provide unique laboratories for
exploring in detail the extreme conditions likely prevailing in the
reionization era.
\end{abstract}
\begin{keywords}
galaxies: individual: SBS\,0335-052E  --- galaxies: starburst ---
galaxies: dwarf --- galaxies: ISM --- galaxies: stellar content 
\end{keywords}
%

\section{Introduction}\label{intro}

The cosmic dawn (6 $\lesssim$ z $\lesssim$ 10) marks a major phase
transition of the universe, during which the ``first light''
[metal-free stars (or the so-called PopIII-stars) and the subsequent
formation of numerous low-mass, extremely metal-poor galaxies]
appeared, putting an end to the dark ages. The details of the
reionization history reflect the nature of these first sources, which
is currently completely unconstrained and the subject of considerable
observational and theoretical efforts
\citep[e.g.,][]{BR13,fbv2014,vbh2017}. In recent years, \emph{Hubble
  Space Telescope} (HST) deep field surveys (e.g., HUDF, CANDELS-deep)
have improved the statistics of galaxies at z $>$ 6. Very soon,
\emph{James Webb Space Telescope} (JWST) will be studying the sources identified at this epoch. At z
$\gtrsim$ 6, however, observing the UV photons (10 nm $<$
  $\lambda$ $<$ 130 nm) gets progressively more
difficult due to the increasing opacity of the intergalactic medium (IGM)
\citep[e.g.,][]{d11}. Even JWST will be able to study in detail (i.e.,
spectroscopically) only the brightest sources \citep[M $>$ 10$^{10}$
M$_{\odot}$;][]{w06}, rather than the more common lower-mass galaxies
that are expected to reionize the universe \citep[e.g.,][]{b15}, for
which only the integrated properties can be derived.

An immediately accessible approach is to identify galaxies at lower
redshifts with properties similar to
galaxies in the very early universe. 
To do this, we need to distinguish the salient observational features that would be associated with such objects.
HeII emission (at $\lambda$1640 \AA~ and
$\lambda$4686 \AA~ in the rest-frame UV and optical ranges,
respectively), observed to be more frequent in high-z galaxies than
locally \citep[e.g.,][]{K11,c13}, is indicative of far harder ionizing radiation than that seen
in nearby systems, as
photons with energy beyond 54.4 eV ($\equiv \lambda \leq$ 228 \AA) are required to twice ionize
He. Star-forming (SF) galaxies with lower metal content tend to
have larger narrow (nebular) HeII line intensities compared to those
with higher metallicities \citep[e.g.,][]{g00}. This agrees
with the expected harder spectral energy distribution (SED) at the
lower metallicities typical in the distant universe. Theoretical arguments
suggest that PopIII-stars and nearly metal-free (Z $\lesssim$
Z$_{\odot}$/100) stars have spectra hard enough to produce many
He$^{+}$-ionizing photons, and so the high-ionization HeII line has been
considered one of the best signatures to single out
candidates for the elusive PopIII-hosting galaxies
\citep[e.g.,][]{TS00,sch03,vhb2015}.
Although recent observations indicate that
high-ionization emission appears to be more usual at the highest
redshifts, the observational signatures of such ionization are not always unambiguous \citep[e.g.,][]{c13,gv15,v16,stark16,mainali17}. Recently, \cite{berg} detected high-ionization narrow emission lines (e.g., CIV1548,1550, HeII1640, OIII]1661,1666) in a metal-deficient lensed galaxy at z $\sim$ 2. They found that the relative emission line strengths can be reproduced with a very high-ionization, low-metallicity starburst with binaries, with the exception of HeII which requires an extra ionization source likely from extreme stellar populations.

The nebular HeII line is the response of the
galaxy gas to the HeII-ionizing continuum ($\lambda \leq$ 228 \AA),
and the hot, massive HeII-ionizing stars, when not obscured, emit most
of their radiation in the FUV ($\lambda \leq$ 2000 \AA). Based
on current facilities, it is not possible to obtain direct
extreme-UV observations of a such star at any redshift. As of today, individual stars
cannot be safely resolved beyond the Local Group, and empirical constraints on
massive stellar models are still limited to the SMC metallicity
\citep[$\sim$ 1/5
Z$_{\odot}$; e.g.,][]{K11,h12,massey13,massey14,garcia14}. Although
significant progress has been made observationally and theoretically, 
modelling massive stellar evolution, in particular at the low
metallicity regime, continues to be challenge 
\citep[e.g.,][]{puls08,tramper11,mui12,langer12,smith14,dori,georgy16,sander17}. 
Ultimately our understanding of metal-poor, hot massive stars remains elusive which propagates to a lack of understanding of the formation of
high-ionization lines as HeII. 
Therefore detailed studies on the origin of nebular
HeII at low redshifts are required to better interpret far-away narrow
HeII-emitters and hence gaining a deeper understanding of the
reionization era. Low metallicity, HeII-emitting local SF galaxies 
provide useful constraints on the little-known SEDs of metal-poor, hot massive
stars and are unique laboratories in which to test stellar population
synthesis models at sub-SMC metallicities \citep[see
e.g.,][]{K08,K13,K15,S17}.  

The goal of this paper is to seek a deeper understanding of the
nebular HeII emission in the nearby galaxy SBS\,0335-052E.
This study is based on \emph{Multi-Unit Spectroscopic Explorer} (MUSE)/VLT
optical integral field spectroscopy and \emph{Chandra}
X-ray observations (see Section~\ref{obs}). SBS\,0335-052E, at a distance of 54 Mpc (NED\footnote{NASA/IPAC Extragalactic Database}), was discovered in the Second Byurakan Survey by \cite{I90}.
It presents an extremely low nebular oxygen abundance of 12+log(O/H)
$\sim$ 7.2-7.3 \citep[$\sim$ 3-4$\%$ Z$_{\odot}$\footnote{assuming a
  solar metallicity 12+log(O/H)$_{\odot}$ = 8.69 (Asplund et al. 2009)}; e.g.,][]{I90,MEL92,I99,P06}, placing it among the most metal-poor SF
galaxies with nebular HeII emission known in the local universe.  The
hard extreme-UV ionizing photons and the very low metallicity gas present in
SBS\,0335-052E are features prevailing in the early universe, which makes this galaxy an excellent low-redshift analog. 

\begin{figure}
\centering
\includegraphics[width=7.5cm]{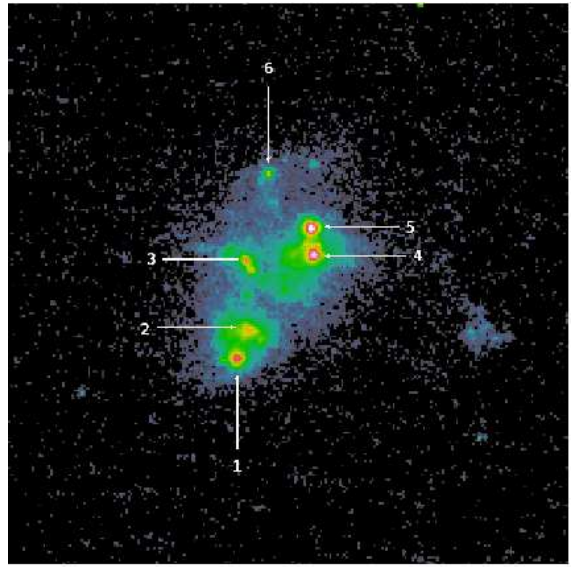}
\caption{HST ACS/F220W archival image of SBS\,0335-052E (HST Proposal
  ID 10575; PI: G.\"Ostlin). The brightest SSCs as identified by Thuan et al. (1997) are labeled. The image is 6'' $\times$ 6''. North is up and east is to the left.}
\label{muse_ifu}
\end{figure}

SBS\,0335-052E undergoes a vigorous starburst and hosts extremely
massive young clusters \citep[see][and references therein]{J09}. All
star formation occurs in six very blue and compact ($<$ 60 pc)
super-star clusters (SSCs) with a mean age of $\sim$ 6 Myrs
\citep[see][]{T97,R08}. In Fig.~\ref{muse_ifu}, we show the \emph{HST/ACS}
UV image of SBS\,0335-052E. In this image
we can see the six brightest SSCs of the galaxy which are found
distributed over a region of $\sim$ 2 arcsec ($\sim$ 520 pc at the
distance of 54 Mpc).

The paper is organized as follows: Section 2 details the observations
and data reduction; Section 3 describes flux measurements and optical emission
line intensity maps, while Section 4 details the spatially resolved
HeII$\lambda$4686-emitting region. The discussion on the origin of the
nebular HeII$\lambda$4686 emission and comparison of models with
observations are covered in Section 5. Finally, Section 6 summarizes
the main conclusions derived from this work.

\section{Observations}\label{obs}

\subsection{Optical Integral field spectroscopic data}

The \emph{Multi-Unit Spectroscopic Explorer} \citep[MUSE;][]{B14}
data of SBS\,0335-052E were obtained in 2015
November 16th and 17th under the ESO program
ID 096.B-0690A (PI: M.Hayes). The observations were carried out in the wide
field mode which provides a field of view of 1 arcmin $\times$
1 arcmin with a sampling
of 0.2 arcsec $\times$ 0.2 arcsec in the wavelength range $\sim$ 4600-9366~\AA.  

We have retrieved the fully reduced data cubes of SBS\,0335-052E from the
ESO archive. The reduction of the raw data used version 1.6.1 of the
MUSE Instrument Pipeline with default parameters, which consists of
the standard procedures of bias subtraction, flat fielding, sky
subtraction, wavelength calibration, and flux calibration. The cube
analysed here is the result of eight exposures of 2840 seconds each,
and the data were taken at airmasses$\sim$ 1.0. The spectral resolving
power is R = $\lambda/\delta\lambda$ = 2988. The PSF FWHM ranges
from $\sim$ 0.6$^{\prime\prime}$ to 0.7$^{\prime\prime}$. 
 We note that these data have
already been independently analysed by \cite{H17}.

\subsection{X-ray Observations} 

SBS\,0335-052E was observed with the front-illuminated CCD array ACIS-I
onboard the \emph{Chandra X-Ray Observatory} on 2000 September 7 for a
total exposure time of 47.6 ks (Observation ID: 796; PI: Thuan).  
Observations were carried out in the Very Faint mode. 
We retrieved level 1 and level 2 processed data from the \emph{Chandra} Data
Center, but actually only the level 1 data were used after reprocessing them
with the \emph{chandra\_repro} tool of the \emph{Chandra} X-Ray Center
software CIAO v4.9 using CALDB 4.7.3.
The data reduction included standard removal of pixel randomization,
cleaning of ACIS background for the Very Faint mode, and bad pixels
masking. 

The event list was further filtered to select good \emph{ASCA} grades.
The data were affected by a few episodes of relatively high background
(above the nominal ACIS-I background rate $\simeq$ 0.20 cnts~s$^{-1}$~chip$^{-1}$ 
in the energy range 0.5-7.0 keV)
that were excised for further data analysis.
The net exposure time was then reduced to 32.5 ks.  
The data analysis was performed using HEASARC FTOOLS and
XSPEC v12.9.1 routines \citep{Arnaud96}.
The astrometry of the X-ray data was refined using X-ray sources in
the FoV spatially coincident with optical sources in the \emph{HST}
archival WFPC2 F791W image.

\section{Optical flux measurements and emission line maps}\label{flux_measurements}

We derived the emission-line fluxes from the $\sim$ 90,000 spaxels\footnote{Individual elements of IFUs are often called ‘spatial
pixels’ (commonly shortened to 'spaxel'); the term is used to
differentiate between a spatial element on the IFU and a pixel on the
detector}  of
the MUSE data cube following the steps described next.  First, we fit the continuum level adjacent to every
emission line for each
spaxel spectrum. This procedure was repeated several times by varying
the local continuum position.  The final continuum level and its
associated uncertainty (or continuum noise) for each line were
adopted to be the average and standard deviation of the repeated
measurements, assuming these follow a normal distribution. After
subtracting the final continuum level from every line, we derive the
line flux by fitting a Gaussian profile to each emission line using
the IDL-based routine MPFIT \citep[][]{ma09}; the peak intensity, the
line width $\sigma$ and the
central wavelength $\lambda_{c}$ for each line are kept as free parameters. In the
case of the H$\alpha$$+$[NII] lines, we fit the three lines
simultaneously keeping the same line width and radial velocity for [NII] and H$\alpha$, and
a nitrogen [NII]$\lambda$6584/[NII]$\lambda$6548 line ratio of 3. 

\begin{figure*}
\center
\includegraphics[width=0.48\textwidth]{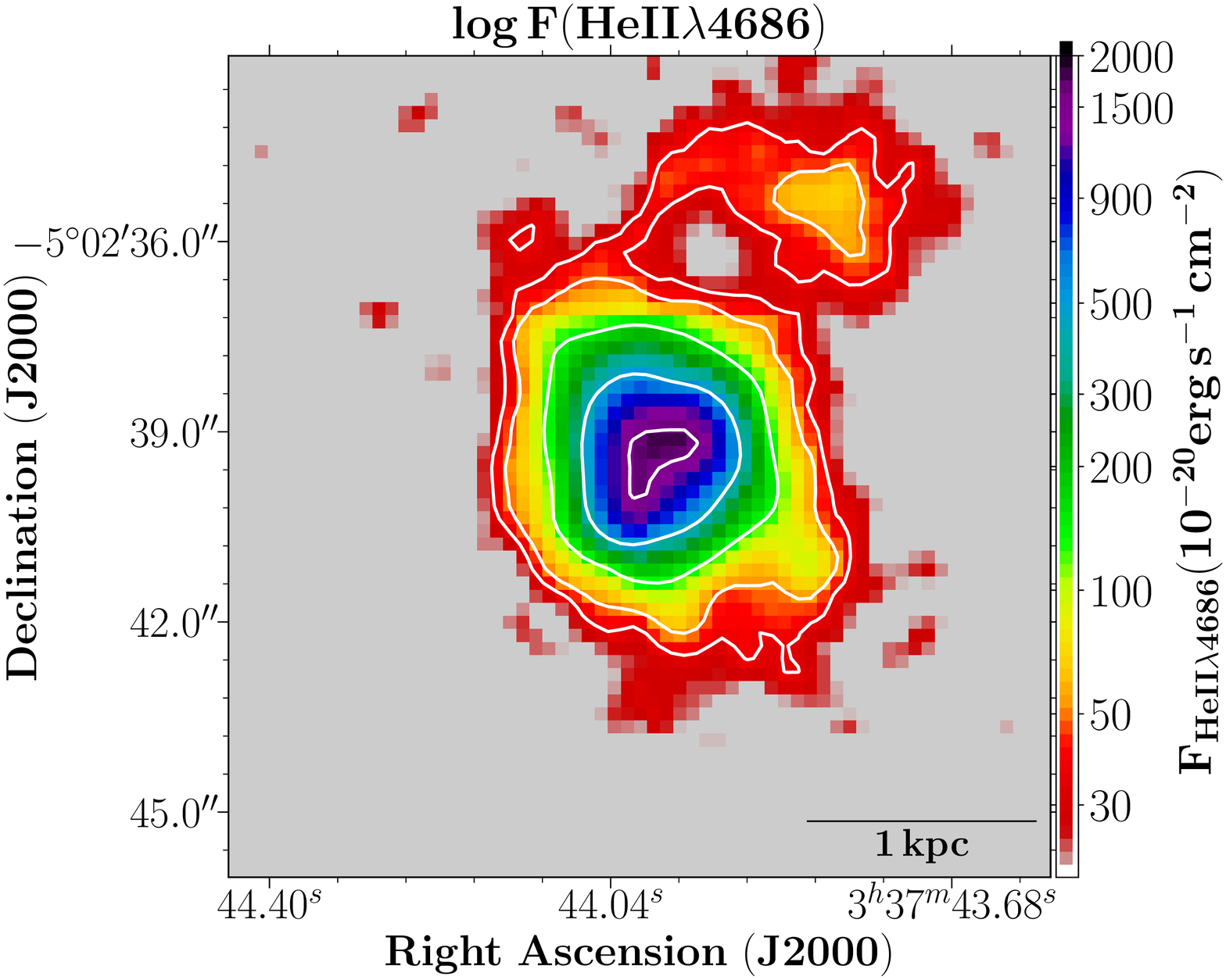}
\includegraphics[width=0.48\textwidth]{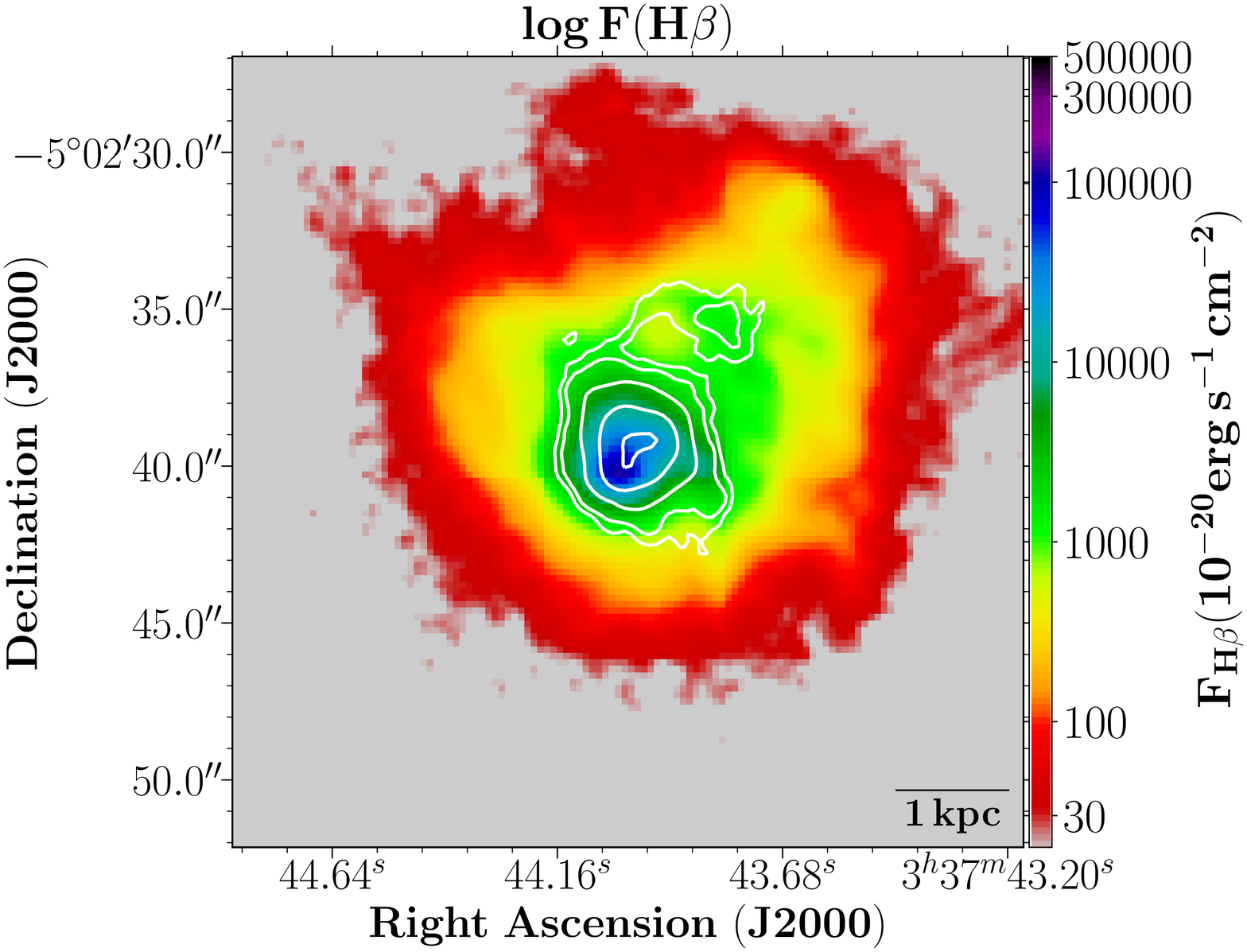}\\
\includegraphics[width=0.48\textwidth]{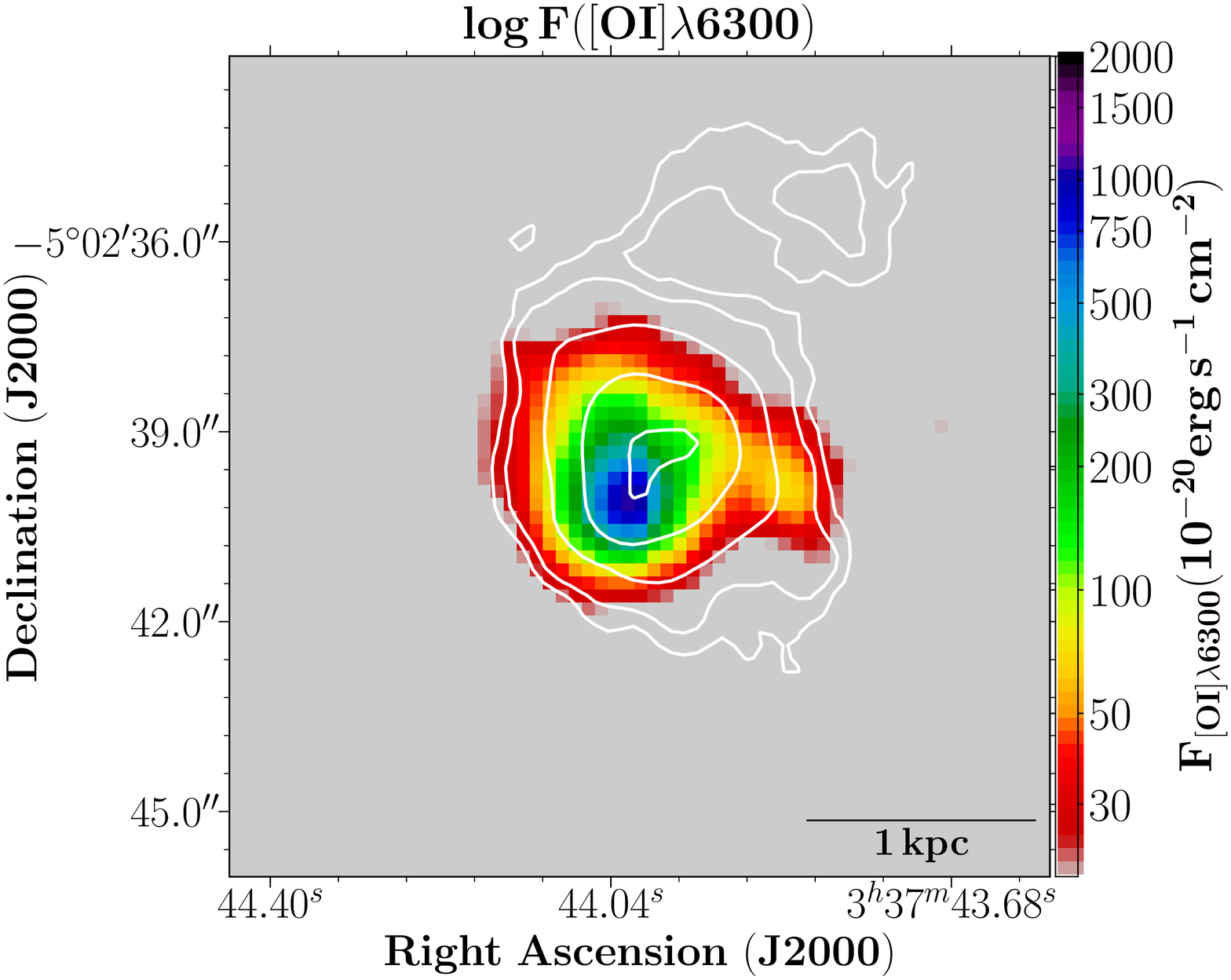}
\includegraphics[width=0.48\textwidth]{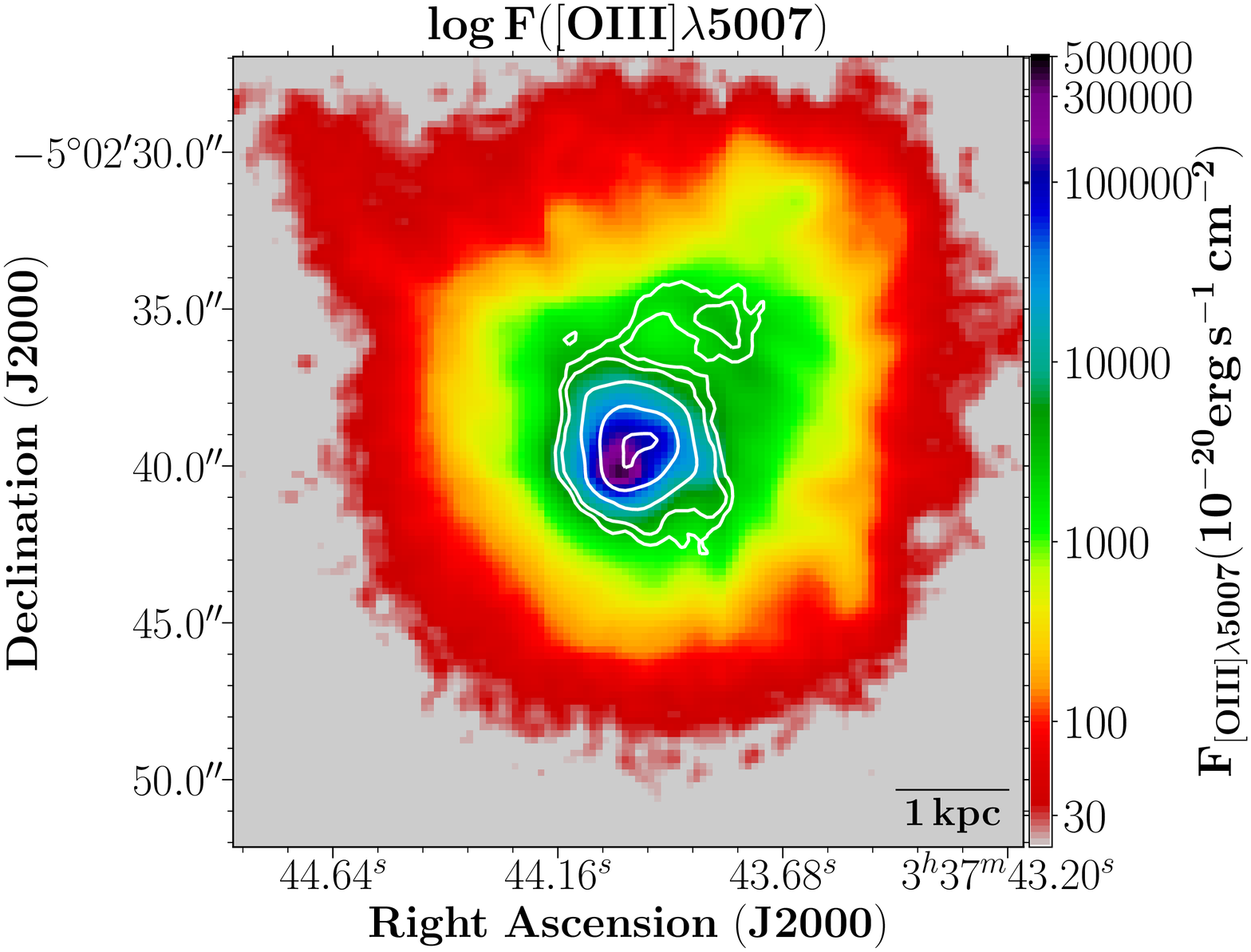}\\
\includegraphics[width=0.48\textwidth]{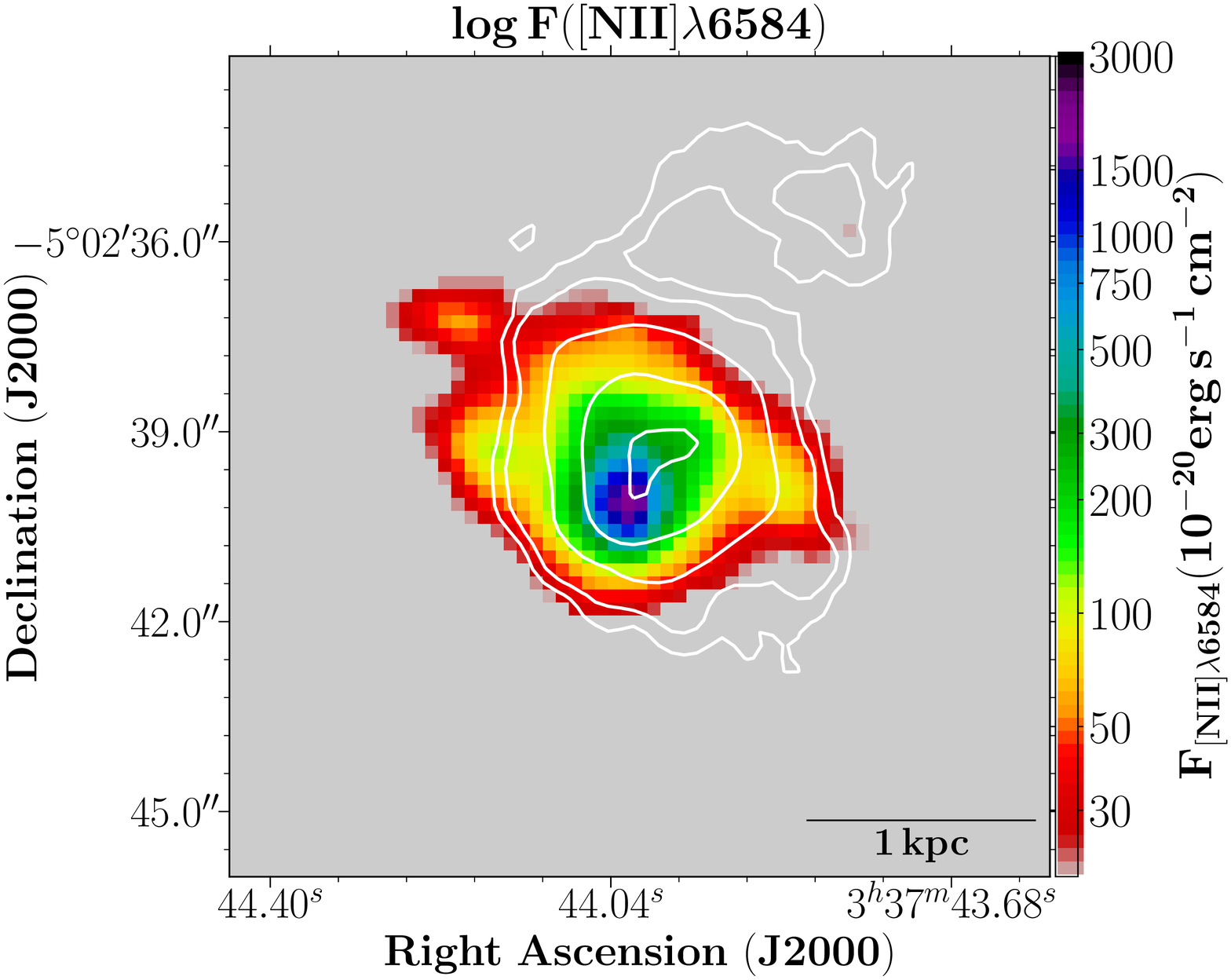}
\includegraphics[width=0.48\textwidth]{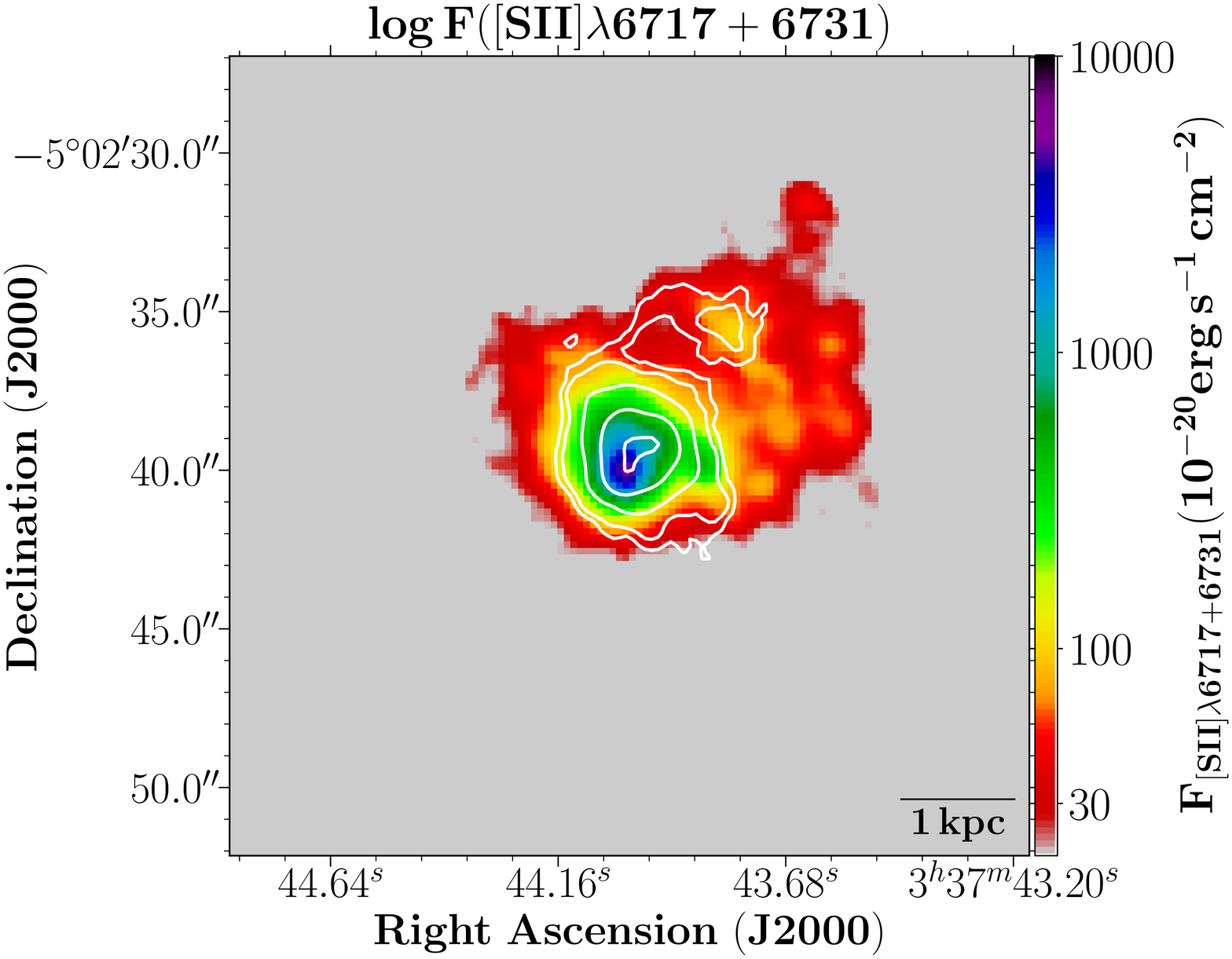}
\caption{Emission line flux maps of SBS\,0335-052E: HeII$\lambda$4686, [OI]$\lambda$6300, [NII]$\lambda$6584, H$\beta$, [OIII]$\lambda$5007, [SII]$\lambda$6717$+$6731. For display purposes, all maps have been smoothed with a $\sigma$ = 1 pixel (0.2'') Gaussian Kernel, and are presented in logarithmic scale. Isocontours of the HeII$\lambda$4686 emission line flux are shown overplotted for reference. The maps for the weakest lines (left column) show a smaller field-of-view in comparison to the brightest ones (right column) for better visualization. North is up and east to the left.}
\label{line_maps} 
\end{figure*}

Using our own IDL scripts, we combine the line fluxes with the
position of the fibres on the sky to create the maps of emission lines
presented in this paper. Fig.~\ref{line_maps} displays the
HeII$\lambda$4686, [OI]$\lambda$6300, [NII]$\lambda$6584, H$\beta$,
[OIII]$\lambda$5007, [SII]$\lambda$6717$+$6731 emission line maps.  As
a guide to the reader, isocontours of the HeII$\lambda$4686 emission line
flux are shown in all maps. The intensity distribution of H$\beta$
and [OIII]$\lambda$5007
are more extended than that of  HeII$\lambda$4686, [OI]$\lambda$6300,
[NII]$\lambda$6584, and [SII]$\lambda$6717$+$6731  because
those lines are among the brightest optical emission lines in
the SBS\,0335-052E spectra. The maps show wisps and/or filaments that surround the
central strong emission region depicting the complex morphology of the
ionized gas in SBS\,0335-052E \citep[see also][]{H17}. 

\section{The spatially resolved HeII$\lambda$4686-emitting region}\label{int}

The presence of narrow HeII$\lambda$4686 emission in SBS\,0335-052E has been
reported before. For instance, \cite{MEL92}, \cite{I97}, and \cite{TI05} based on
long-slit spectroscopy, show the narrow HeII$\lambda$4686 line in the spectrum of
SBS\,0335-052E.  \cite{I06} first produced a 
HeII$\lambda$4686 spectral map of SBS\,0335-052E using the VLT/GIRAFFE spectrograph.
However, the superior spatial resolution of MUSE provides much more
detailed emission-line flux maps, and therefore allows us to perform a broader
study on the origin of the HeII ionization in SBS\,0335-052E, a
subject still under debate not only for this galaxy but for other
several local HeII-emitting SF systems
\citep[e.g.,][]{G91,I06,K11,sb12,K13,K15}. We will discuss this point further in Section~\ref{discu}.

\begin{figure*}
\center
\includegraphics[width=0.95\textwidth]{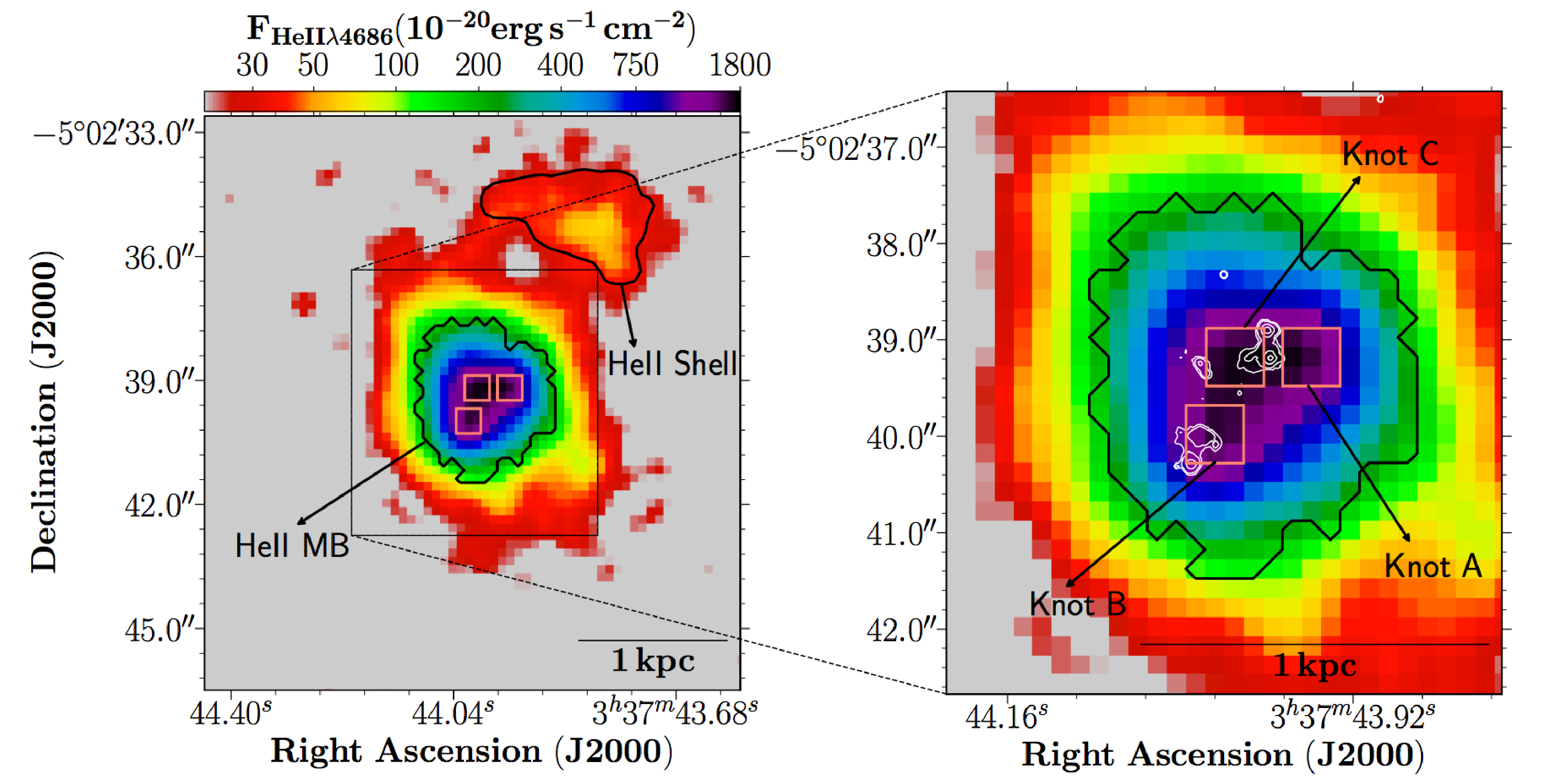}\\
\includegraphics[width=0.48\textwidth]{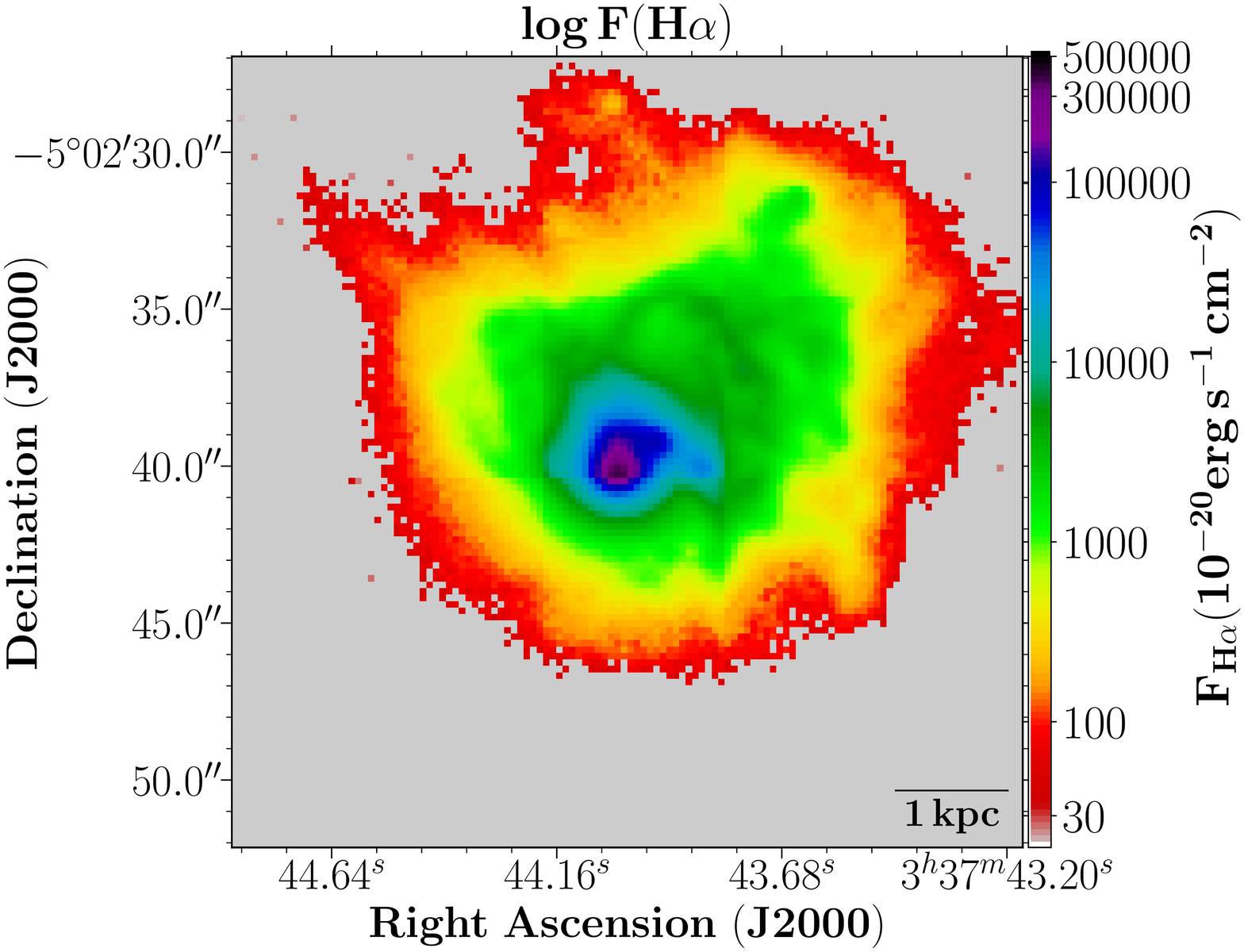}
\caption{{\it Top row}: The HeII$\lambda$4686 map in two different
  scales showing the boundaries of the areas that we use to create the
  integrated spectra of the HeII-emitting regions: HeII shell, HeII main body, and HeII knots A,B,C (see text and table 1 for details); in the top-right panel the white lines mark the location of the brightest SSCs of SBS\,0335-052E (see Fig.1). {\it Bottom row}: The H$\alpha$ map showing only the spaxels with H$\alpha$ S/N (per pixel) $>$ 10. North is up and east to the left.}
\label{fig_regions} 
\end{figure*}

Figures~\ref{line_maps} and \ref{fig_regions} reveal a highly extended
HeII$\lambda$4686-emitting region from our IFS data \citep[see
also][]{H17}. The spatial distribution of HeII is quite different from
that of the other emission lines (see Fig.\ref{line_maps}) The HeII
emission consists of a roundish oval shape component with a diameter
of $\sim$ 5 arcsec ($\sim$ 1.3 kpc at the distance of 54 Mpc) over the
galaxy core, and a shell-like structure (called here ``HeII shell'')
which are connected to each other. The HeII-emitting zone extends out
to distances $\gtrsim$ 6 arcsec ($\sim$ 1.5 kpc) from the youngest SSCs \citep[SSCs \#1
and \#2; see e.g.,][]{R08}, and presents three peaks spatially
displaced from the brightest stellar clusters (see {\it top-right}
panel from Fig.~\ref{fig_regions}).

From our data, we checked that the FWHM of the HeII$\lambda$4686 line
is comparable to that of the other nebular emission lines (e.g.,
H$\beta$, [OIII]$\lambda$5007). The measured values of the mean
($\mu$) and standard deviation ($\sigma$) for the
FWHM(HeII)/FWHM(H$\beta$) and FWHM(HeII)/FWHM([OIII]$\lambda$5007
ratios are $\mu$ = 1.13 ($\sigma$ = 0.09) and $\mu$ = 1.10 ($\sigma$ =
0.10), respectively. The narrow line profile for the HeII$\lambda$4686
emission and its spatial extent are evidence of its nebular nature.

Based on the spatial distribution of the emission in HeII we created
the 1D spectra for several regions by summing the flux in the spaxels
within each of the corresponding areas displayed in
Fig.~\ref{fig_regions}.  The spectra of the knots centered on the
three peaks seen in the HeII line map are named HeII
Knots A, B and C; the area of their aperture extraction is 0.36
arcsec$^{2}$ each. The 'HeII shell' integrated
zone covers $\sim$ 9.7 arcsec$^{2}$.  We have also obtained the 1D
spectrum for which we term HeII main body ('HeII-MB') by adding all
HeII-emitting spaxels with HeII S/N $>$ 10; thus exclusively the
emission of the brighter HeII ionized gas has been integrated in the
HeII MB spectrum.
Finally, we obtained for the first time the integrated spectrum of
SBS\,0335-052E. To do so we have integrated the flux in all the spaxels
for which H$\alpha$ S/N (per spaxel) $>$ 10; this corresponds to an area of
$\sim$ 340 arcsec$^{2}$ ($\sim$ 23 kpc$^{2}$), enclosing basically all the nebular emission across our MUSE
FOV (see the H$\alpha$ map in Fig.~\ref{fig_regions}). 

\begin{figure*} 
\includegraphics[width=8.0cm,clip]{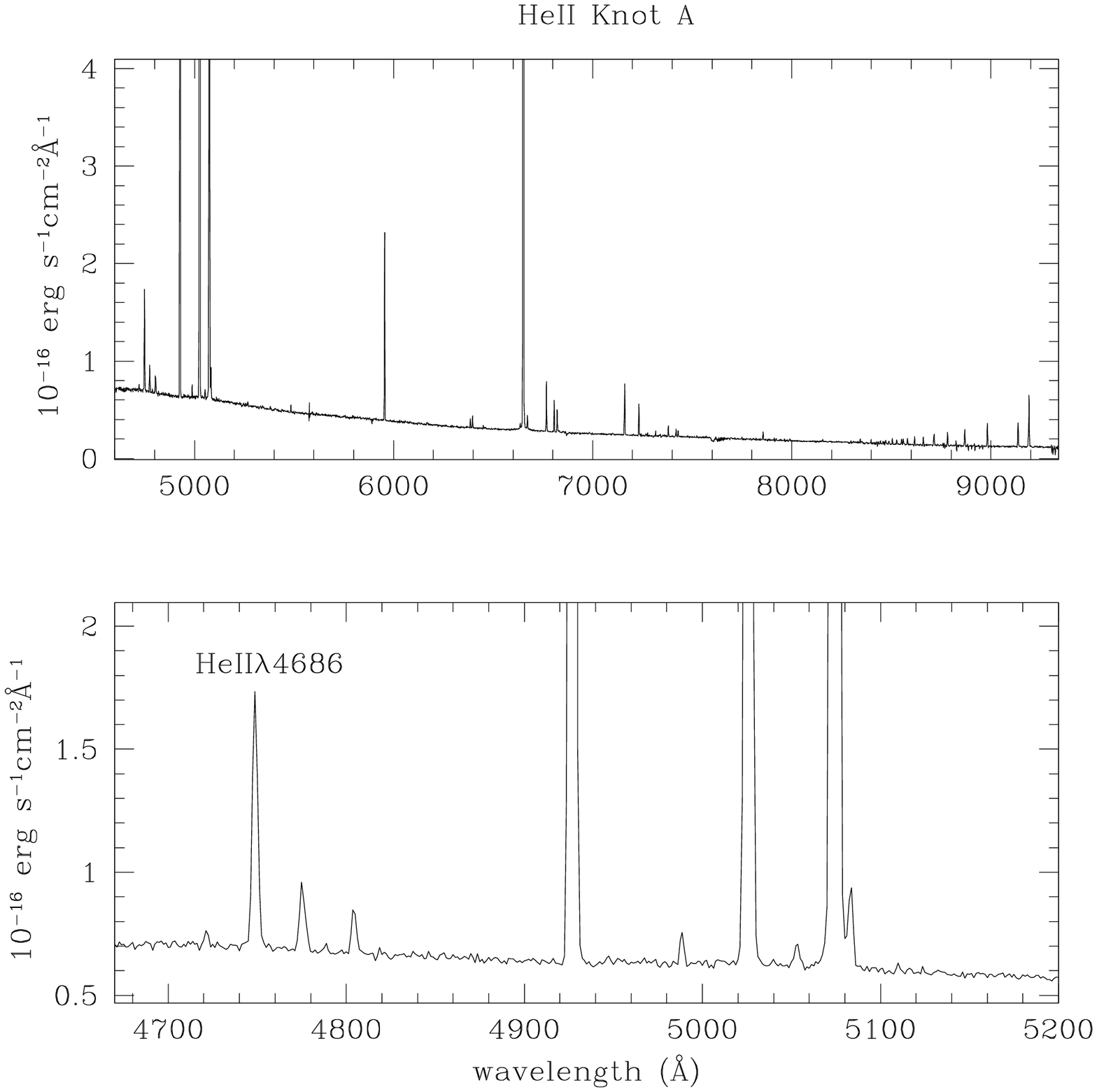} 
\includegraphics[width=8.0cm,clip]{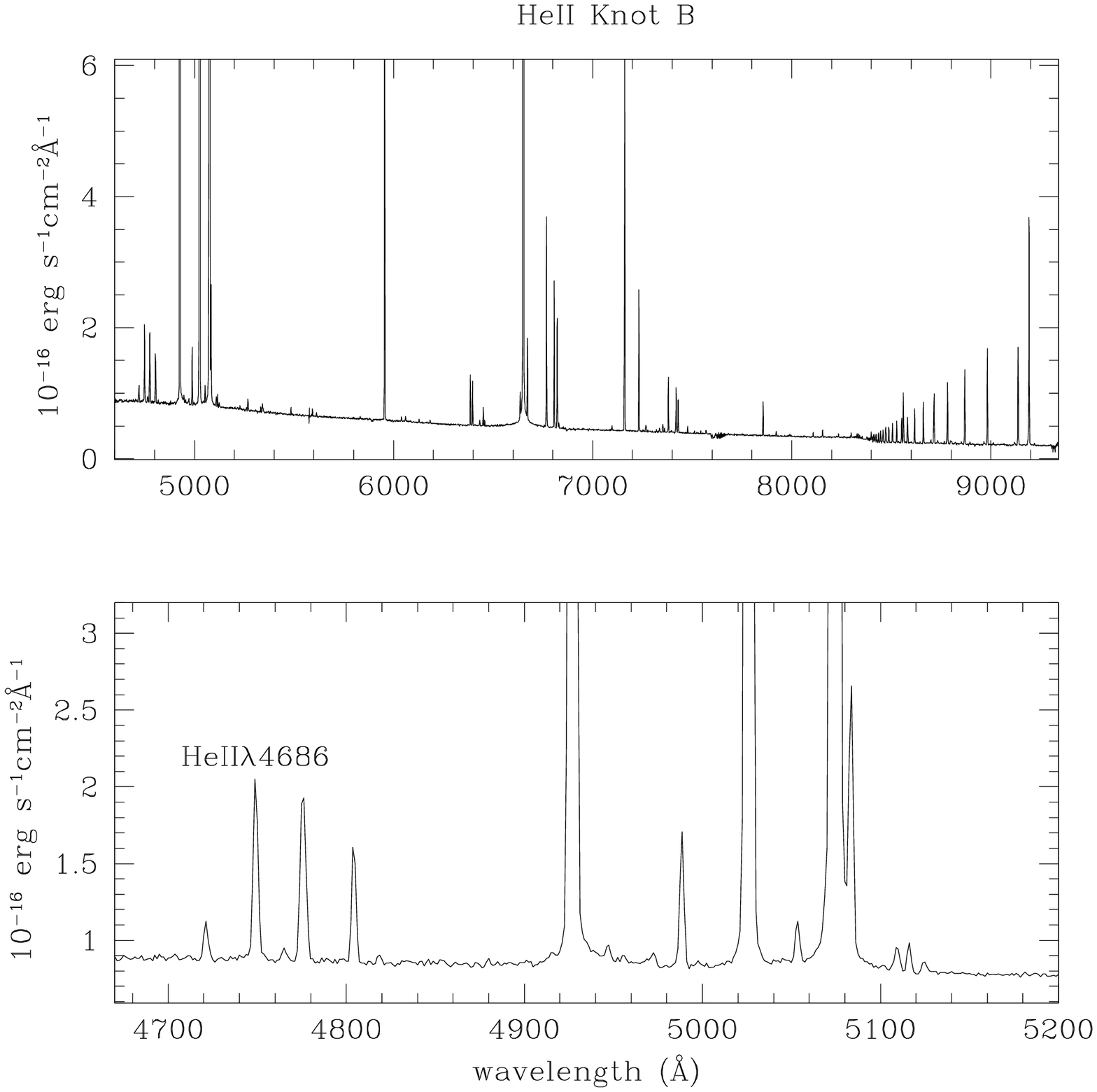} \\
\includegraphics[width=8.0cm,clip]{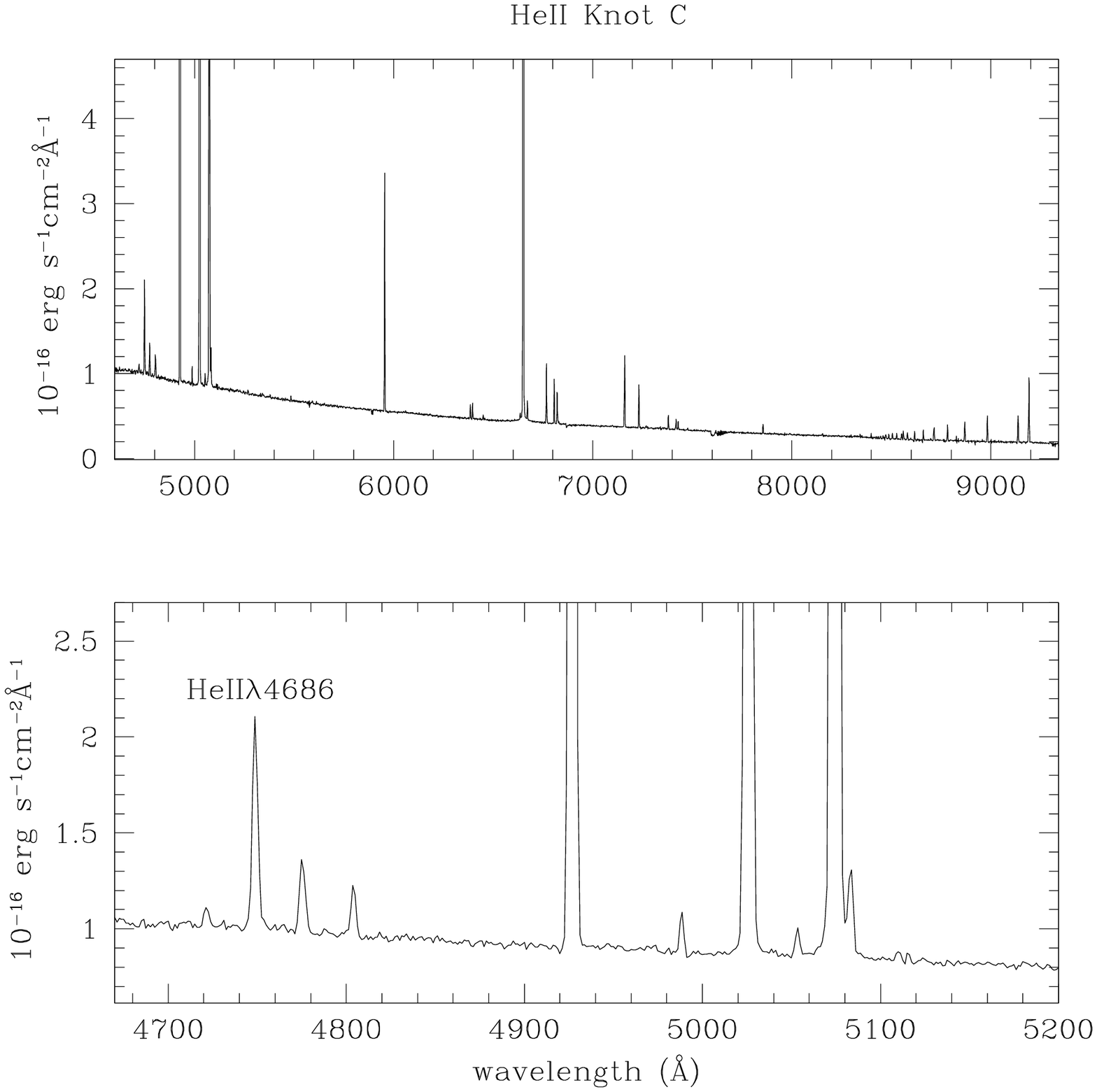} 
\includegraphics[width=8.0cm,clip]{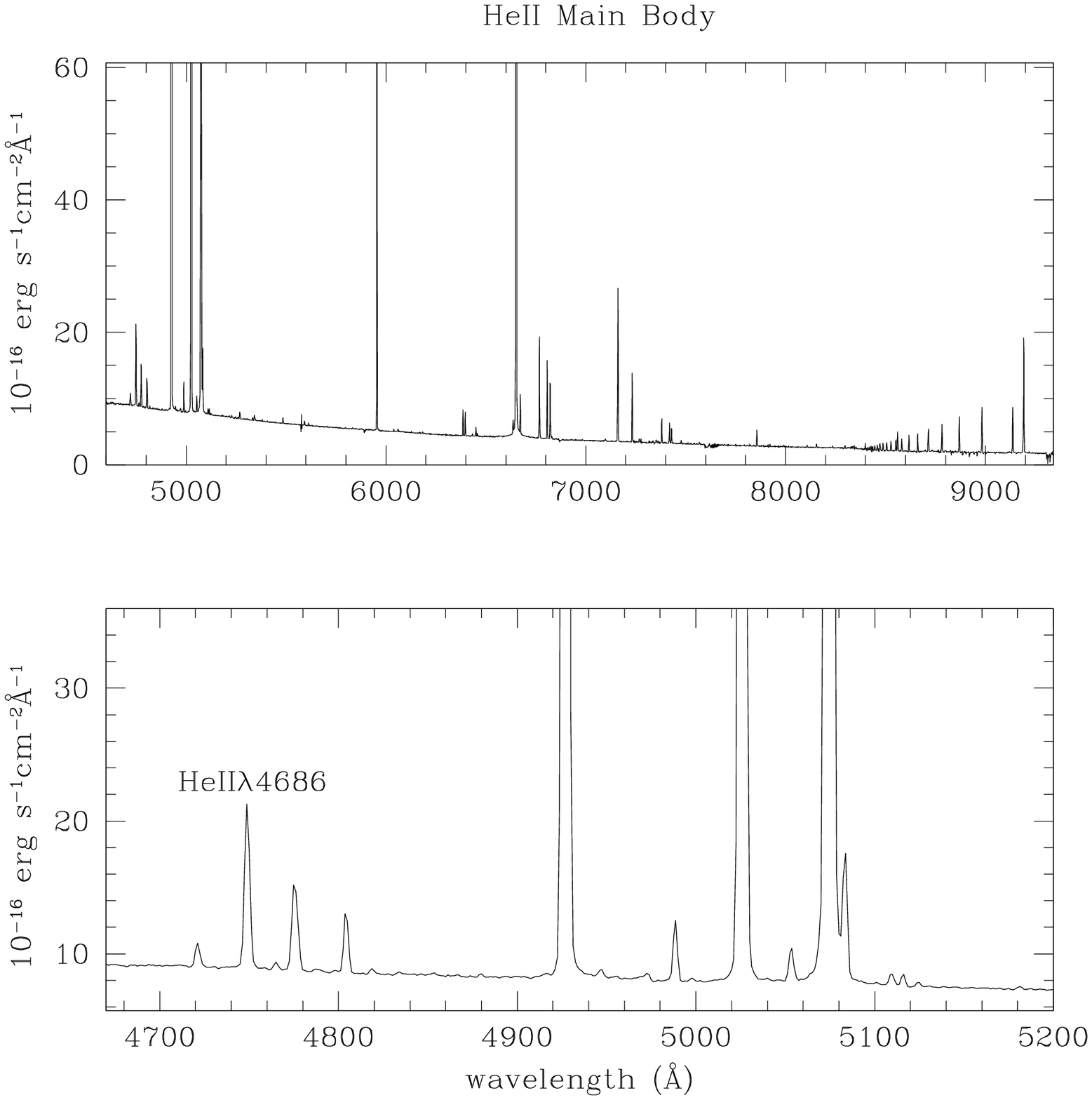} \\
\vspace{-0.25cm}
\includegraphics[width=8.0cm,clip]{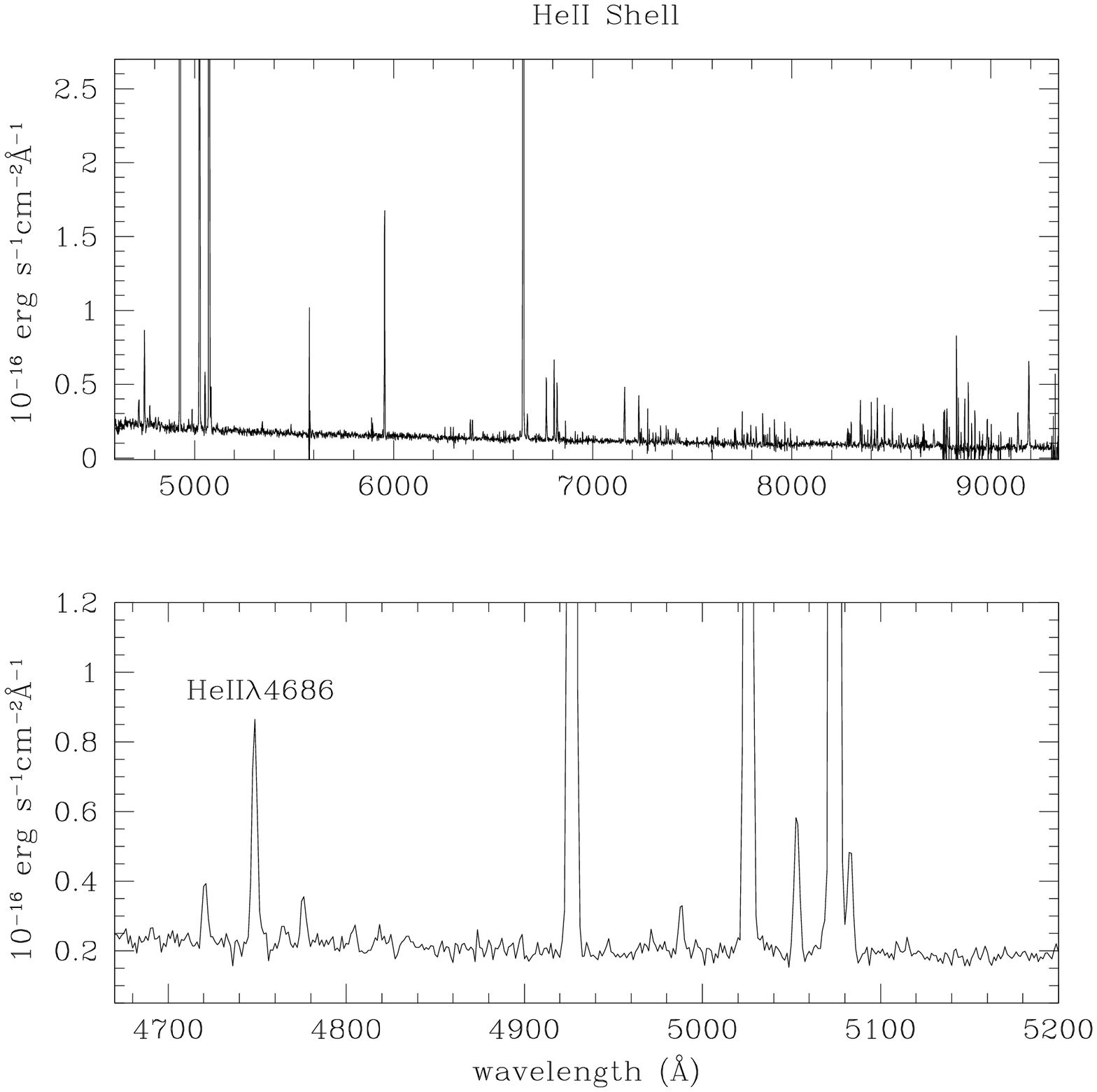}
\includegraphics[width=8.0cm,clip]{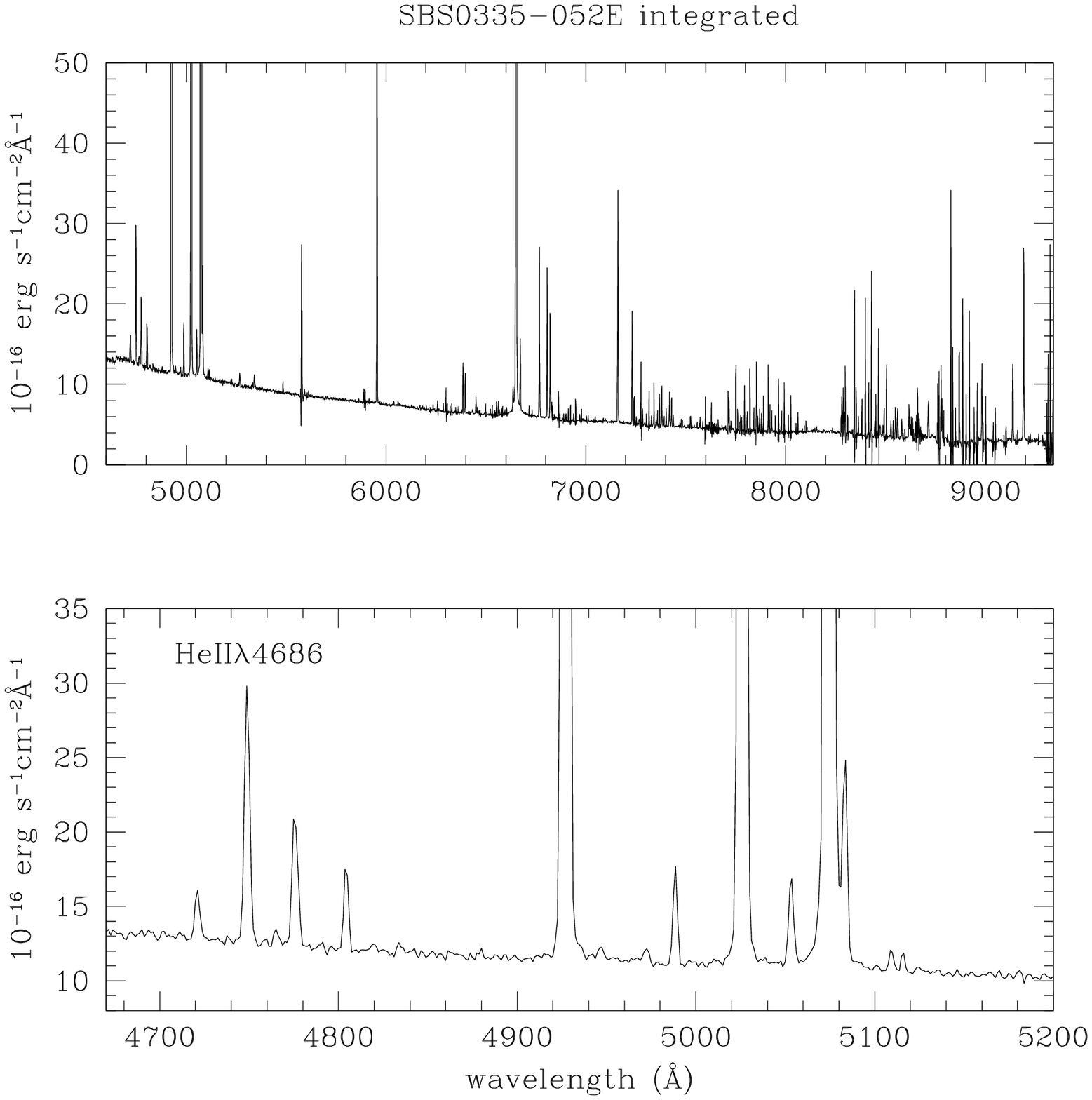}
\vspace{-0.5cm}
\caption{Flux-calibrated 1D spectra of the selected HeII-emitting regions and the integrated spectrum of SBS\,0335-052E (see the
text for details). A zoom of the wavelength range $\sim$
4670-5200 \AA~ is also displayed. The spectra are in units of 10$^{-16}$ erg s$^{-1}$ cm$^{-2}$ \AA$^{-1}$.}
\label{1d_heii_knots} 
\end{figure*}

Fig.\ref{1d_heii_knots} presents the 1D spectra for the aforementioned
regions of SBS\,0335-052E. We measured the emission line fluxes of the
1D spectra with the SPLOT routine in IRAF\footnote{IRAF is distributed
  by the National Optical Astronomical Observatories, which are
  operated by the Association of Universities for Research in
  Astronomy, Inc., under cooperative agreement with the National
  Science Foundation.} by integrating all the line flux between two
points given by the position of a local continuum. The continuum level
is estimated by visually placing the graphics cursor at both sides of
each line.  This process was repeated several times for each emission
line by varying the continuum position. We take the mean and the
standard deviation of the repeated measurements as the final flux of
each line and its associated uncertainty, respectively. The reddening
coefficient, c(H$\beta$), corresponding to each summed spectra was
computed from the ratio of the measured-to-theoretical
H$\alpha$/H$\beta$ assuming the reddening law of \cite{cardelli89},
and case B recombination with electron temperature $T_{e}$ =
2$\times$10$^{4}$ K and electron density $n_{e}$ = 100 cm$^{-3}$ \citep[e.g.,][]{P06} which give an intrinsic value of H$\alpha$/H$\beta$= 2.75
\citep[][]{OF06}. Reddening-corrected line intensities, normalized to
H$\beta$, along with physical properties obtained from the 1D spectra
are shown in Table~\ref{table_regions}.

The HeII ionizing photon flux, Q(HeII), can be derived from the
  reddening-corrected L$_{HeII\lambda4686}$, using the relation
  Q(HeII) = L$_{HeII\lambda4686}$/[j($\lambda$4686)/$\alpha_{B}$(HeII)]
  (Osterbrock \& Ferland 2006).  Making use of the integrated
  spectrum, and assuming case B recombination and an electron
  temperature $T_{e}$ = 2$\times$10$^{4}$ K \citep[e.g.,][]{P06,I06},
  we computed for the first time the total HeII ionizing photon flux,
  Q(HeII)$_{Int}$ = 3.17 $\times$ 10$^{51}$ photons s$^{-1}$, in
  SBS\,0335-052E. Q(HeII)$_{Int}$ is a relevant quantity in order to
  study the origin of the nebular HeII emission \citep[see][and see also
  Section~\ref{discu}]{K15}. Using the same methodology, we also
  computed Q(HeII) for the other selected regions of SBS\,0335-052E
  mentioned above (see Table~\ref{table_regions}).  From
  Table~\ref{table_regions} we can see that the HeII main body and the
  HeII shell, together, produce a total of $\sim$ 2.38 $\times$
  10$^{51}$ He$^{+}$-ionizing photons s$^{-1}$ which represents $\sim$
  75$\%$ of the integrated Q(HeII)$_{Int}$. This indicates that the
  contribution to the Q(HeII)$_{Int}$ from the most external and faint
  HeII emission from the SBS\,0335-052E core is not negligible ($\sim$ 25$\%$). This
  highlights the importance of high-spatial resolution IFS for our analysis, which
  have allowed us to collect all HeII emission free from aperture
  effect corrections required in single-fiber or long-slit
  spectroscopic observations.

\begin{table*}
\caption{De-reddened emission line-fluxes relative to 1000$\cdot$I(H$\beta$) 
and physical properties from selected regions} 
\label{table_regions}
 \centering 
\begin{minipage}{15.0cm}
\centering
\begin{tabular}{lcccccc}
\hline\hline 
Wavelength & HeII Knot A  & HeII Knot B  & HeII knot C & HeII ``shell'' & HeII-MB$^{a}$ & Integrated$^{b}$ \\  \hline
4658 [Fe~III]  & 2.4 $\pm$ 0.4  &  2.60 $\pm$ 0.10  & 3.5 $\pm$ 0.4  & 10.5 $\pm$ 0.5  & 3.40 $\pm$ 0.11 & 4.47 $\pm$ 0.26  \\ 
4686 He~II    & 60.8 $\pm$ 0.6  & 14.20 $\pm$ 0.10  & 46.0 $\pm$ 0.7 & 39 $\pm$ 1  & 27.14 $\pm$ 0.13 & 26.08 $\pm$ 0.32  \\
4714 [Ar~IV]+He~I  & 16.3  $\pm$ 0.5 & 16.10 $\pm$ 0.20  & 16.8 $\pm$ 0.5 & 7 $\pm$ 1  & 16.49 $\pm$ 0.11 & 14.93 $\pm$ 0.29  \\
4740 [Ar~IV]  & 9.54  $\pm$ 0.36  & 8.70 $\pm$ 0.10   & 9.8 $\pm$ 0.5 & ---   & 8.99 $\pm$ 0.10 & 7.79 $\pm$ 0.18   \\
4861 H$\beta$ & 1000 $\pm$ 1  & 1000 $\pm$ 4  & 1000 $\pm$ 1  & 1000 $\pm$ 1   & 1000 $\pm$ 1 & 1000 $\pm$ 4  \\
4959 [O~III]  & 1052 $\pm$ 1  & 1099 $\pm$ 3  & 1171  $\pm$ 1  & 774 $\pm$ 1   & 1094 $\pm$ 1 & 1006 $\pm$ 4   \\
5007 [O~III]  & 3128 $\pm$ 2  & 3231 $\pm$ 9  & 3487  $\pm$ 2  & 2336 $\pm$ 2  & 3271 $\pm$ 2 & 3002 $\pm$ 8  \\
5876 HeI    & 87.29 $\pm$ 0.20 & 128.95 $\pm$ 0.38 & 93.99 $\pm$ 0.28 & 89 $\pm$ 1 & 105.75 $\pm$ 0.25 & 101.34 $\pm$ 0.33 \\
6300 [O~I]    & 3.70 $\pm$ 0.20 & 8.78 $\pm$ 0.06 & 5.15 $\pm$ 0.12  & 7.5  $\pm$ 0.4   & 7.17 $\pm$ 0.04 & 8.04 $\pm$ 0.15   \\
6312 [S~III]  & 5.20 $\pm$ 0.20 & 7.23 $\pm$ 0.06  & 5.60 $\pm$ 0.13   & 7.3  $\pm$ 0.4  & 6.36 $\pm$ 0.05 & 6.09 $\pm$ 0.23  \\
6563 H$\alpha$ & 2750 $\pm$ 3 & 2665 $\pm$ 9  &  2750 $\pm$ 7   & 2750  $\pm$ 3   & 2750 $\pm$ 13 & 2750 $\pm$ 10  \\
6584 [N~II]   & 4.90 $\pm$ 0.30 & 12.00 $\pm$ 0.30 & 5.90  $\pm$ 0.26    & 10 $\pm$ 1   & 9.30 $\pm$ 0.04 & 10 $\pm$ 1   \\ 
6678 HeI    &  22.50 $\pm$ 0.10 & 33.70 $\pm$ 0.10 & 23.36  $\pm$ 0.21  &  24 $\pm$ 1  & 27.02 $\pm$ 0.11 & 26.10 $\pm$ 0.15  \\   
6717 [S~II]  & 13.50 $\pm$ 0.10 & 23.30 $\pm$ 0.10 & 16.43  $\pm$ 0.11   &  28.66 $\pm$ 0.38   & 20.48 $\pm$ 0.11 & 22.79 $\pm$ 0.39  \\  
6731 [S~II]  & 11.20 $\pm$ 0.10 & 19.80 $\pm$ 0.10 & 13.02  $\pm$ 0.13  &  22.2 $\pm$ 0.4  & 17.0 $\pm$ 0.5 & 17.1 $\pm$ 0.6   \\ \hline 
c(H$\beta$)    & 0.22  & 0.00    & 0.20   & 0.13  & 0.10 & 0.10  \\
-EW(H$\alpha$) (\AA)  & 656   & 1319    & 600   & 1400   & 1081  & 1030  \\
-EW(H$\beta$) (\AA)  & 97   & 300     & 96   & 305   & 187 & 199  \\
F(H$\alpha$)$^{c}$ & 2.85$\times10^{-14}$ & 6.98$\times10^{-14}$ & 3.76$\times10^{-14}$ & 2.20$\times10^{-14}$ & 5.42$\times10^{-13}$ & 7.95$\times10^{-13}$ \\
F(H$\beta$)$^{c}$ & 1.03$\times10^{-14}$ & 2.62$\times10^{-14}$ &1.37$\times10^{-14}$ & 8.02$\times10^{-15}$  & 1.97$\times10^{-13}$ & 2.89$\times10^{-13}$ \\
F(He~II)$^{c}$  & 6.29 $\times10^{-16}$ & 3.73$\times10^{-16}$ &6.30$\times10^{-16}$ & 3.14$\times10^{-16}$  & 5.35$\times10^{-15}$ & 7.54$\times10^{-15}$ \\ 
L(H$\beta$)(erg s$^{-1}$)  &3.61$\times10^{39}$  &9.13$\times10^{39}$  &4.77$\times10^{39}$ &2.04$\times10^{39}$  &6.88$\times10^{40}$ & 1.01$\times10^{41}$ \\
L(He~II)(erg s$^{-1}$)  &2.19$\times10^{38}$  &1.30$\times10^{38}$  &2.20$\times10^{38}$ &8.94$\times10^{37}$  &1.87$\times10^{39}$ &2.63$\times10^{39}$  \\
Q(He~II)(photon s$^{-1}$)$^{d}$ &2.64$\times10^{50}$  &1.56$\times10^{50}$ &2.65$\times10^{50}$  &1.32$\times10^{50}$  &2.25$\times10^{51}$ & 3.17$\times10^{51}$  \\  
Q(H)(photon s$^{-1}$)$^{e}$ & 1.38$\times10^{52}$ & 3.39$\times10^{52}$ & 1.83$\times10^{52}$ & 1.07$\times10^{52}$ &2.63$\times10^{53}$ &3.86$\times10^{53}$   \\ \hline
log ([O~I]6300/H$\alpha$) & -2.88  & -2.48  & -2.73  & -2.57   & -2.59 & -2.53  \\
log ([N~II]6584/H$\alpha$) & -2.74 & -2.35  & -2.67 & -2.43  & -2.47 & -2.44  \\
log ([S~II]6717+31/H$\alpha$) & -2.05  & -1.79 & -1.97  & -1.73  & -1.86 & -1.84  \\
log ([O~III]5007/H$\beta$) & 0.49 & 0.51  & 0.54  & 0.37  & 0.51 & 0.48  \\
$n_{\rm e}$([S~II])(cm$^{-3}$)  & 264  & 318 & 168 & 121  & 267 & 60  \\  
\hline
\end{tabular}
\end{minipage}
\begin{flushleft}
(a) ``HeII main body'' (HeII-MB) spectrum obtained by summing all HeII-emitting spaxels with He~II S/N $>$ 10.\\ 
(b) SBS\,0335-052E integrated spectrum created by adding all spaxels with H$\alpha$ S/N $>$ 10.\\
(c) H$\beta$, H$\alpha$ and He~II extinction-corrected fluxes in units of erg s$^{-1}$ cm$^{-2}$.\\
(d) Number of ionizing photons shortward of the He$^{+}$ edge. \\
(e) Number of ionizing photons shortward of the H$^{0}$ edge.
\end{flushleft}
\end{table*}

\section{On the origin of the HeII$\lambda$4686 emission and ionizing sources}\label{discu}

Despite numerous attempts to explain the origin of nebular HeII
emission in nearby and distant SF systems, there are many cases that
are currently
inconclusive \citep[e.g.,][]{g00,K11,c13,gv15,K15,S17}. The most
popular scenarios for producing nebular HeII emission involve X-ray
sources, shocks and hot stellar ionizing continua
\citep[e.g.,][]{G91,d96,c02,sch03,TI05,sb12,K15}. In the following we discuss each of these
possible HeII ionizing sources in
SBS\,0335-052E.

\subsection{X-ray emission}

\begin{figure*}
\includegraphics[width=7.5cm]{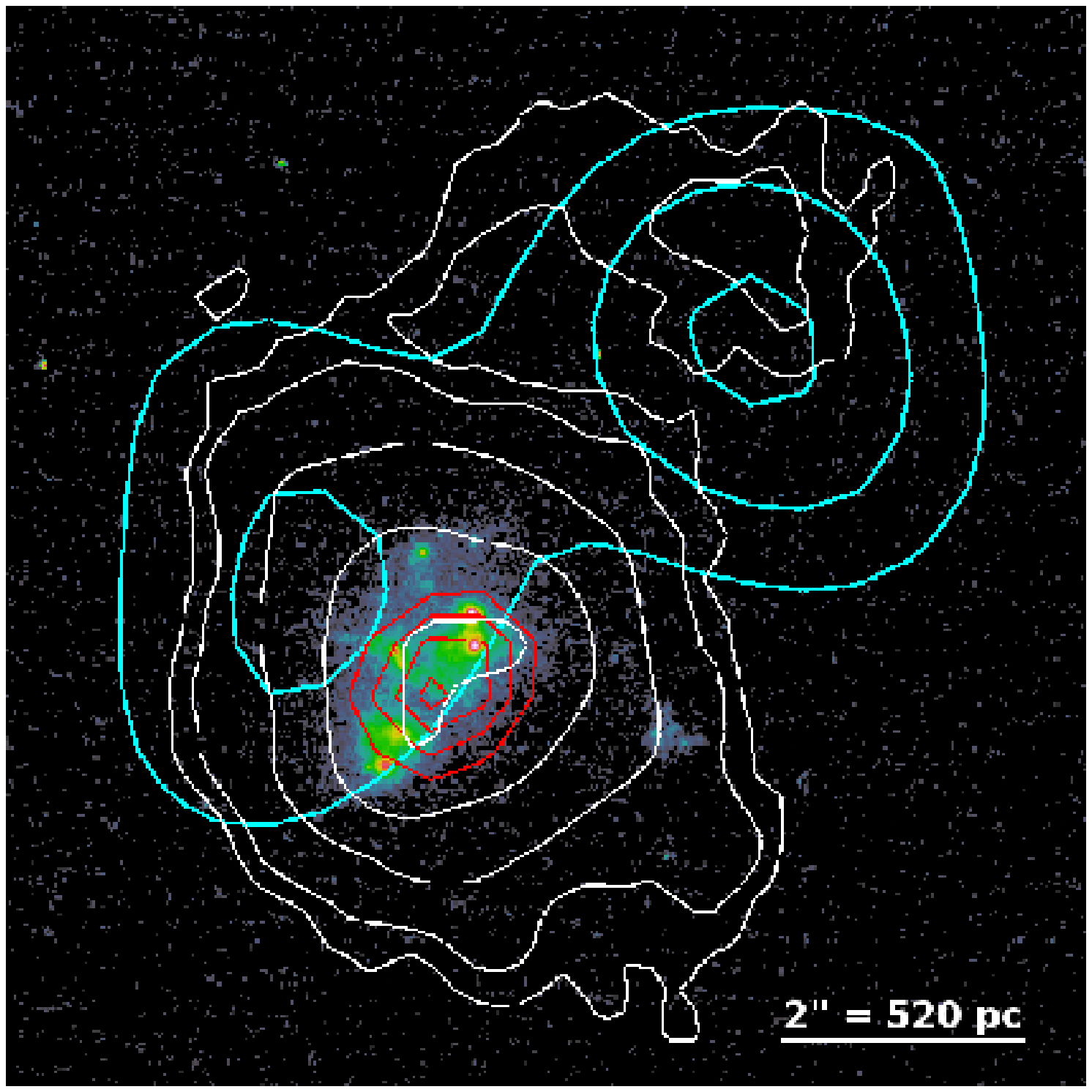}
\includegraphics[width=7.5cm]{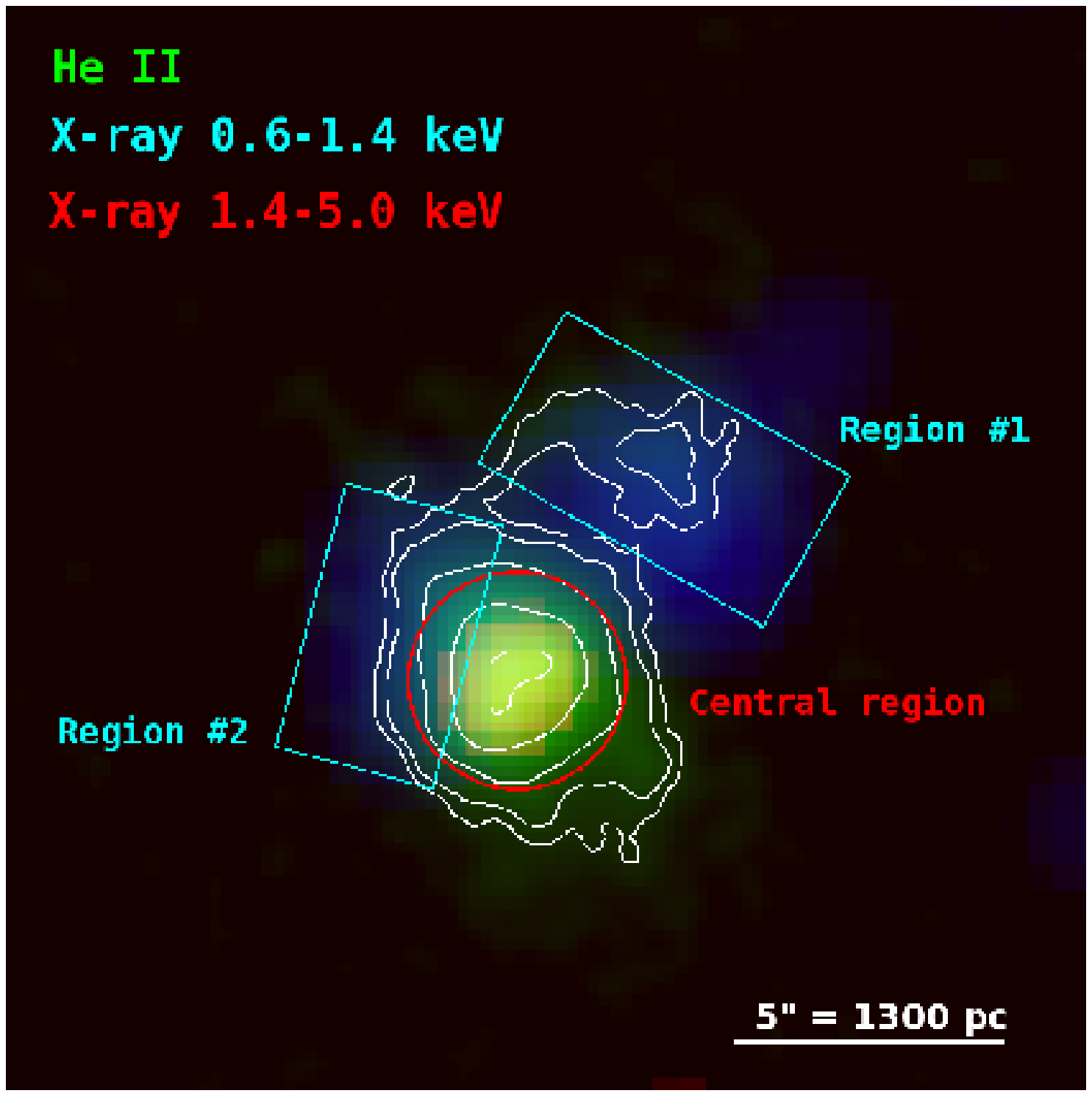}
\caption{Left panel: Soft (cyan lines) and hard (red lines) X-ray contours overlaid on the HST ACS/F220W image of SBS\,0335-052E. Right panel: Colour composite image of SBS\,0335-052E in three bandpasses (blue = soft X-rays, red = hard X-rays, green = HeII$\lambda$4686 flux image).  In both panels, the white curves represent the isocontours of the HeII$\lambda$4686 emission line flux. North is up and east is to the left.}
\label{xray}
\end{figure*}

A preliminary inspection of the \emph{Chandra} observations of
SBS\,0335-052E confirms the detection of faint X-ray emission 
as reported by \cite{T04}.  
To investigate the correspondence between the spatial distribution of the
X-ray, UV, and optical emissions in SBS\,0335-052E, we made adaptively
smoothed X-ray images in soft (0.6--1.4 keV) and hard (1.4--5.0 keV) bands
with spatial resolution from 1\farcs2 for bright point sources up to 
1\farcs8 for weak diffuse emission.  
These X-ray images are compared to the \emph{HST} ACS/F220W
and HeII images in Fig.~\ref{xray}.

The comparison between the X-ray emission and the UV image reveals a source of hard X-ray
emission (red contours in Fig.~\ref{xray}-\emph{left}) spatially coincident with the brightest
SSCs at the core of SBS\,0335-052E.
At the spatial resolution of the X-ray images, the source is 
basically unresolved, but for a slight ellongation along the
Southeast-Northwest direction.
\cite{T04} associated this source with SSC \#2, 
although some additional emission was associated with SSCs
\#3, \#4, and \#5 (the SSCs are labeled in Fig.~\ref{muse_ifu}).
Our analysis cannot provide a definitive association between this
X-ray source and any of the SSCs; although at lower resolution, we can see that its peak is indeed
closer to SSC \#2 (see red contours in Fig.~\ref{xray}-\emph{left}). However, 
the distribution of the soft X-ray emission is utterly different from that 
of the hard X-ray emission, with one faint blob of diffuse emission towards 
the Northeast of the core of SBS\,0335-052E and another one brighter towards
the Northwest (cyan contours in Fig.~\ref{xray}-\emph{left}).  
The comparison of the soft X-ray and HeII images reveals that
the latter source of soft X-ray emission is confined by the HeII 
Northwest shell (see Fig.~\ref{xray}-\emph{left} and \emph{right}).  

According to the X-ray images presented above, we have extracted X-ray
spectra from the apertures labeled ``central region'', ``region \#1'',
and ``region \#2'', as shown in Fig.~\ref{xray}-\emph{right}, corresponding to
the central hard point source, and to the Northwest and Northeast sources
of soft diffuse emission, respectively.
The count numbers of these spectra are too low to allow 
detailed spectral fits.  
Indeed, the spectral analysis of the X-ray emission from the hard source
presented by \cite{T04}  could constrain neither the emission
model (power-law or optically-thin plasma), nor its parameters.
Assuming a hydrogen column density $N_{\rm H}=1\times10^{22}$
cm$^{-2}$, consistent with that used by \cite{T04},
and a metallicity 3$\%$ $Z_\odot$ \citep[e.g.,][]{I99,P06,I06}, we find that the emission
in this hard source is consistent with an optically-thin plasma
emission model with a temperature of 3 keV.
For comparison, Thuan et al.\ (2004) suggested plasma
temperatures $\sim$3 keV, and certainly in excess of
1.2 keV.
As for the diffuse X-ray emission, the same metallicity of 3$\%$
$Z_\odot$ was assumed, but the hydrogen column density was reduced
to $N_{\rm H}=5\times10^{20}$ cm$^{-2}$ to be consistent with the
optical extinction $c({\rm H}\beta)=0.13$ derived for ``region \#1'' and ``region \#2'' from the MUSE
data.
With these assumptions, the (even fainter) X-ray spectrum of the diffuse
X-ray emission is found to be consistent with an optically-thin plasma
emission model at a temperature of 1 keV.

The intrinsic X-ray luminosities of the central hard point-like source 
and the Northwest soft diffuse emission in the 0.3-5.0 keV band are 
4$\times$10$^{39}$ and 3$\times$10$^{38}$ erg~s$^{-1}$, respectively.
These X-ray models (spectral shapes, luminosities, absorptions) have been
convolved with the energy-dependent cross section of He$^+$ to estimate
the effective X-ray ionizing power.
We find $\log$ Q(HeII) to be 36.2 for the central hard source 
and 35.4 for the Northwest source confined within the 
HeII shell.
Clearly, these numbers show that the X-ray emission from the different
sources found in SBS\,0335-052E cannot provide the ionizing flux responsible
for its HeII emission (see Table~\ref{table_regions}).  
Moreover, the different models to describe the X-ray emission from
SBS\,0335-052E (for instance, a power-law) also result in similarly low
values of $\log$ Q(HeII) that are well below the required value to ionize
HeII at the level observed in this galaxy.  

\subsection{A 2D view of the gas excitation from MUSE}

\begin{figure} 
  \centering
\includegraphics[bb=27 23 596 468,width=8.0cm,clip]{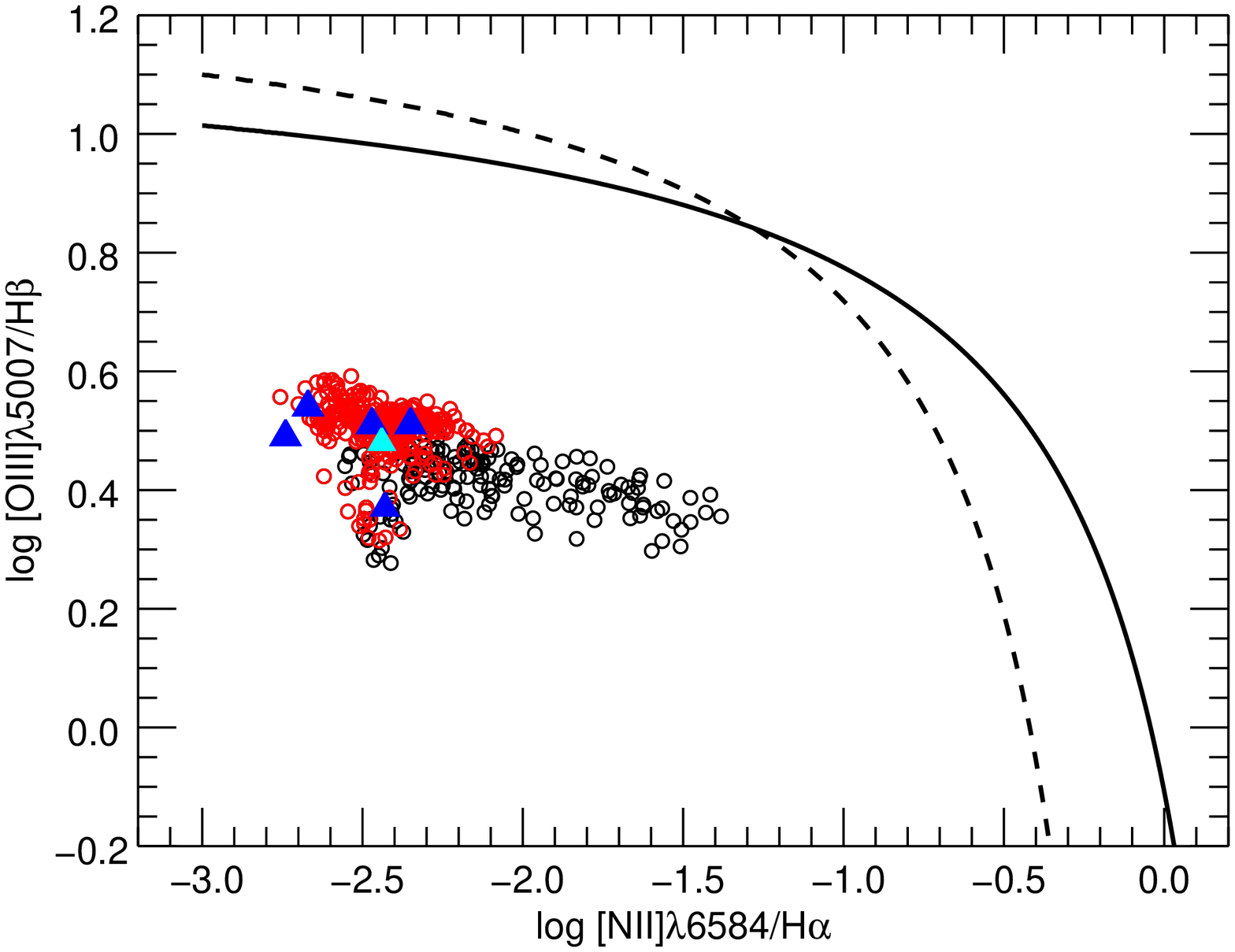} 
\includegraphics[bb=27 23 596 468,width=8.0cm,clip]{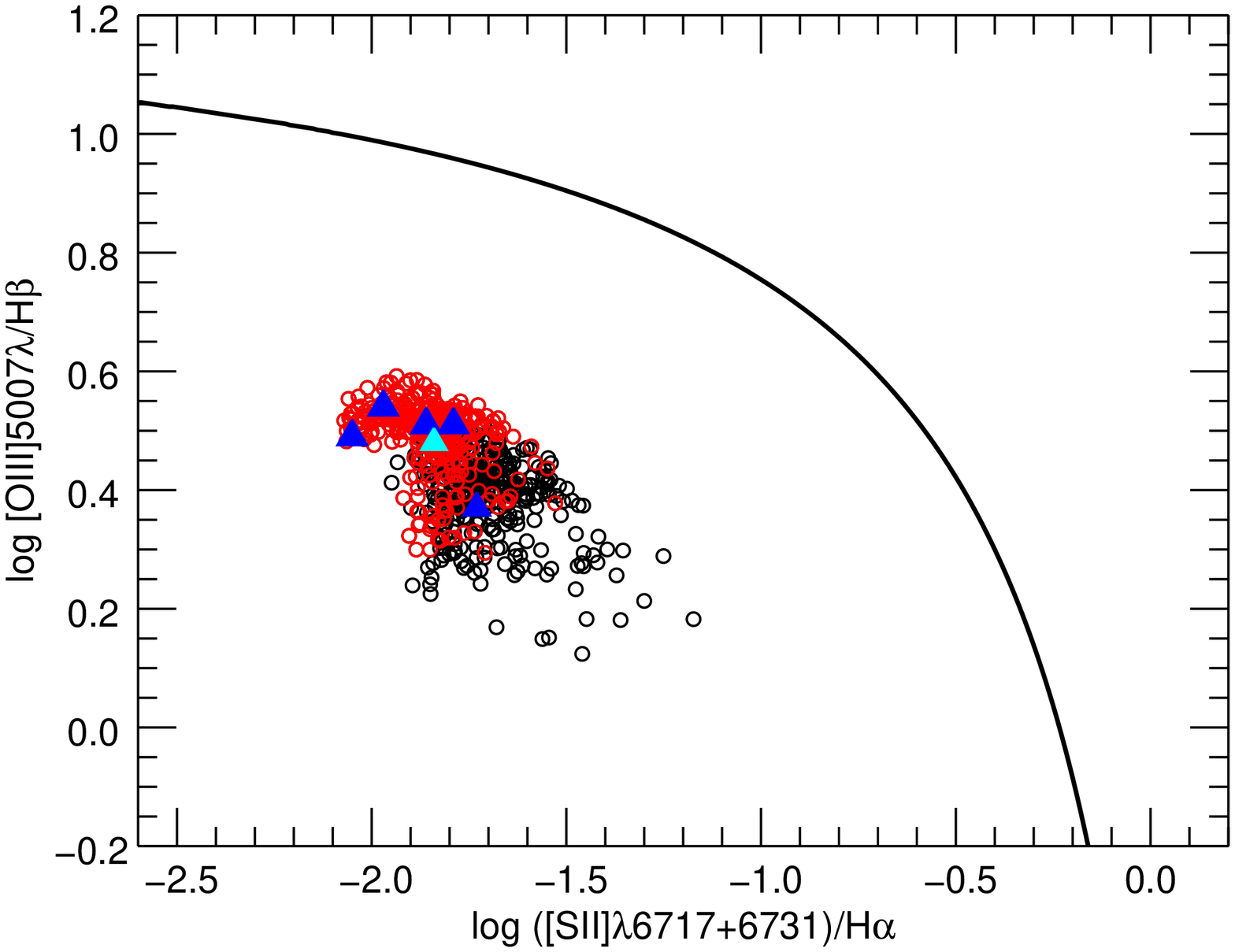} 
\includegraphics[bb=27 23 596 468,width=8.0cm,clip]{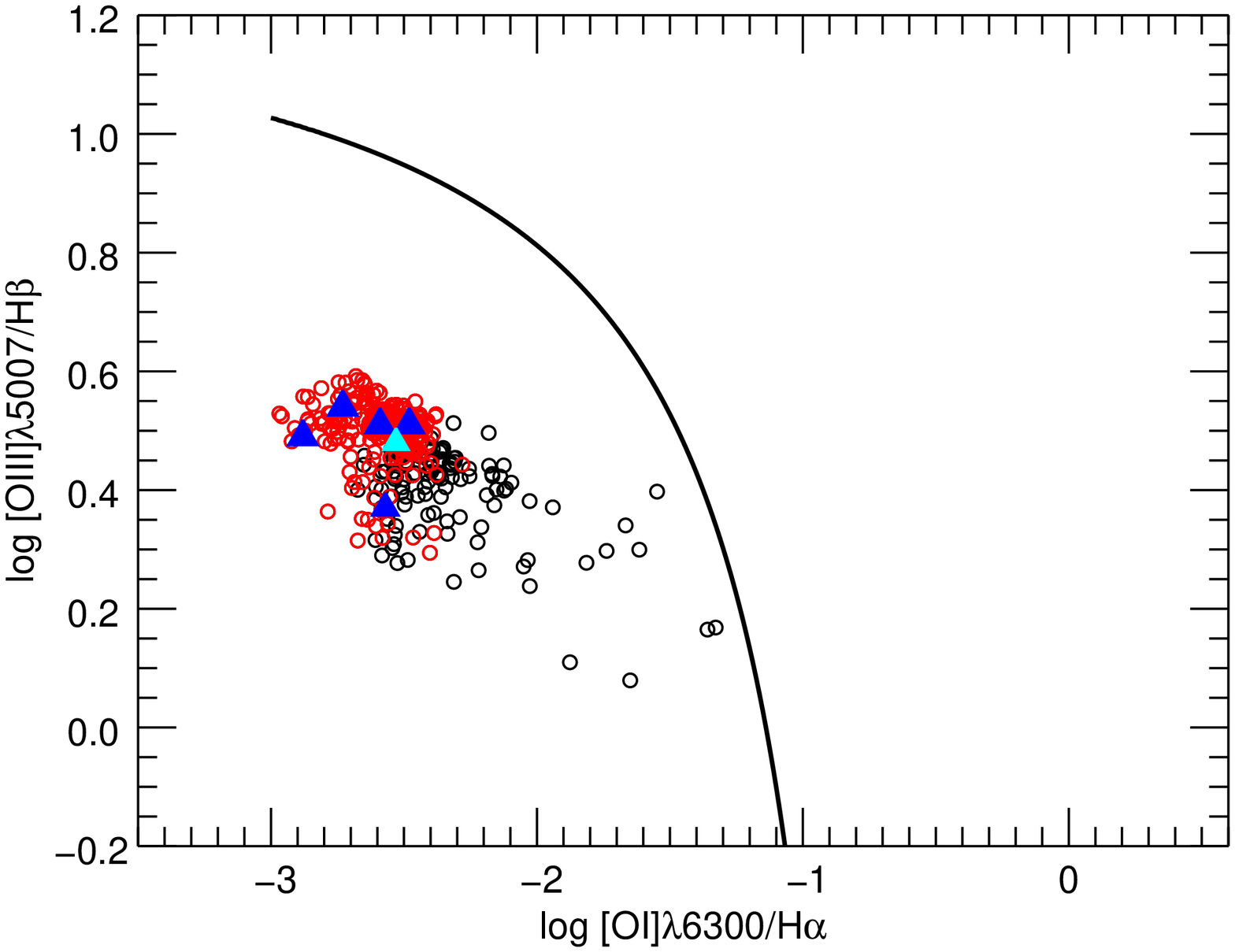}
\vspace{-0.3cm}
\caption{BPT diagnostic diagrams for SBS\,0335-052E. From top to bottom: log ([O{\sc iii}]$\lambda$5007/H$\beta$) vs. log ([N{\sc
    ii}]$\lambda$6584/H$\alpha$), log ([O{\sc
    iii}]$\lambda$5007/H$\beta$) vs. log ([S{\sc
    ii}]$\lambda$6731,6717/H$\alpha$) and log ([O{\sc
    iii}]$\lambda$5007/H$\beta$) vs. log ([O{\sc
    i}]$\lambda$6300/H$\alpha$). Only fluxes with S/N $\geq$ 5 are shown. Open circles
correspond to individual spaxels from the data cube; red circles mark
the individual HeII-emitting spaxels.  Overlaid as triangles are the
line ratios measured from the 1D spectra of selected galaxy regions;
the cyan triangle  corresponds to the total integrated spectrum of
SBS\,0335-052E (see Section~\ref{int} and
Table~\ref{table_regions}). The black solid curve (in
all three panels) is the theoretical maximum starburst model from
\protect\cite{kew01}, devised to isolate objects whose emission line ratios can be accounted for by the photoionization by massive stars (below and to the left of the curve) from those where some other source of ionization is required. The black-dashed curve in the [NII]/H$\alpha$ diagram represents the demarcation between SF galaxies (below and to the left of the curve) and AGNs defined by \protect\cite{K03}.}
\label{bpt} 
\end{figure}

\begin{figure*}
\center
\includegraphics[width=0.43\textwidth]{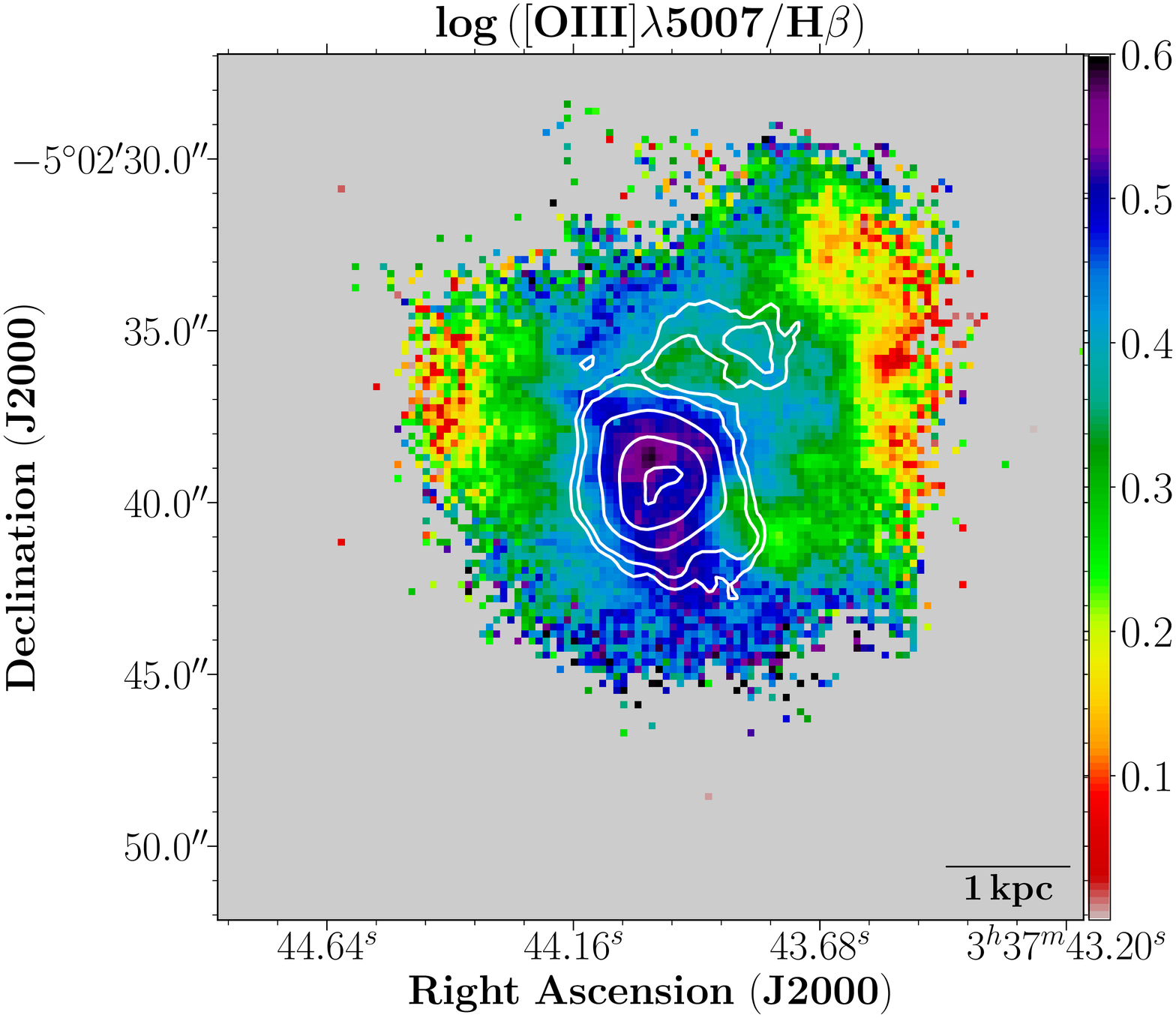}
\includegraphics[width=0.45\textwidth]{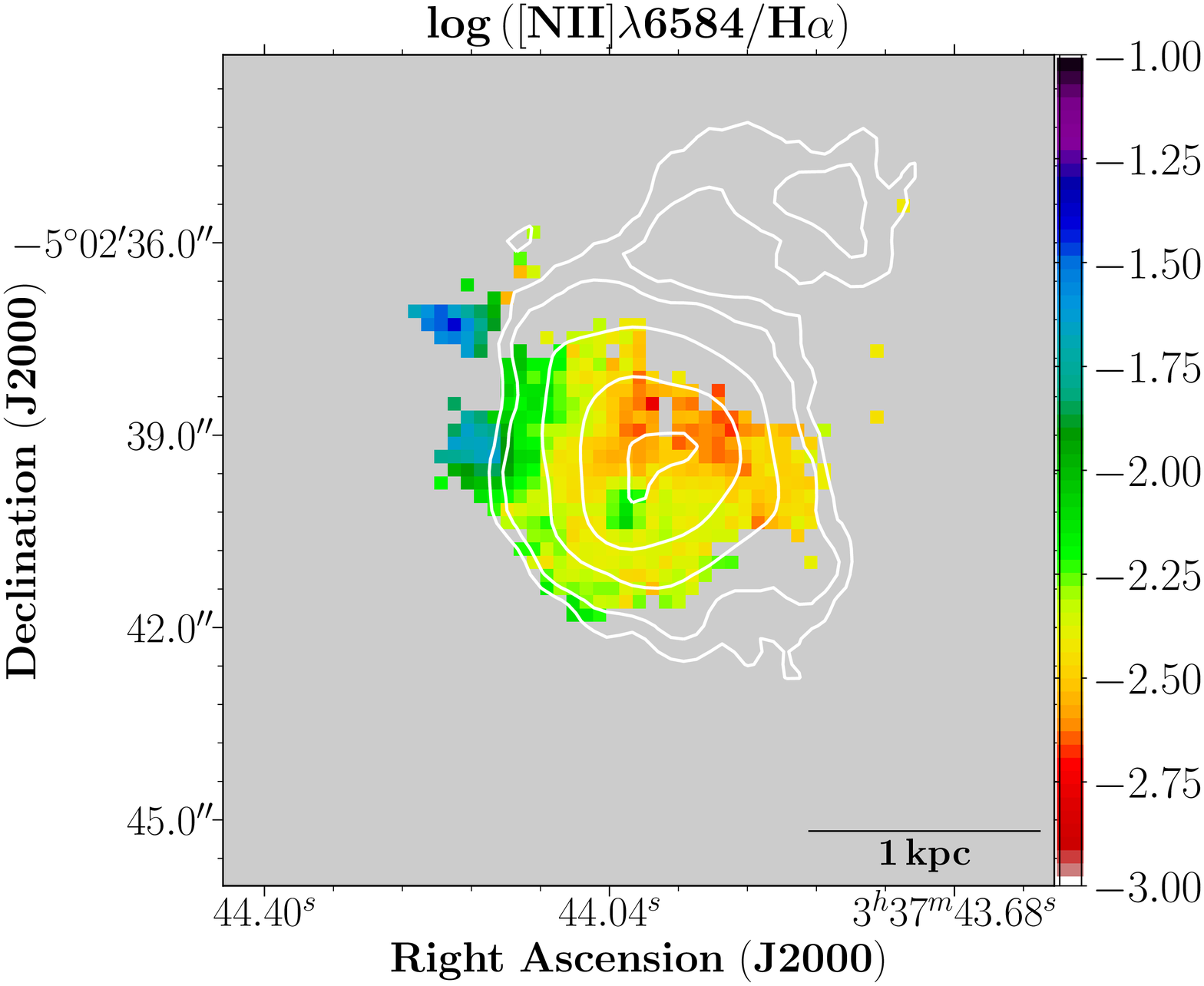}\\
\includegraphics[width=0.45\textwidth]{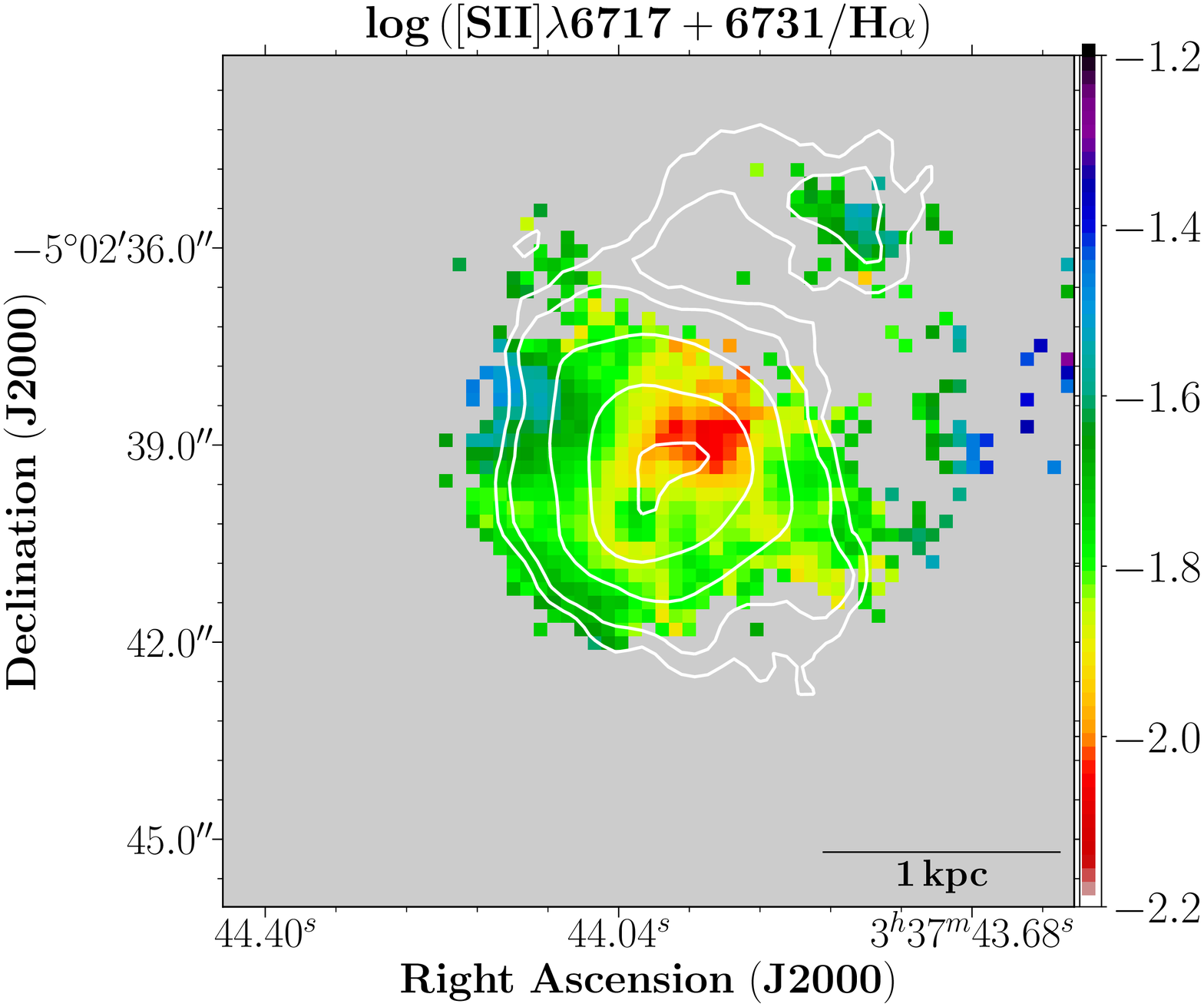}
\includegraphics[width=0.45\textwidth]{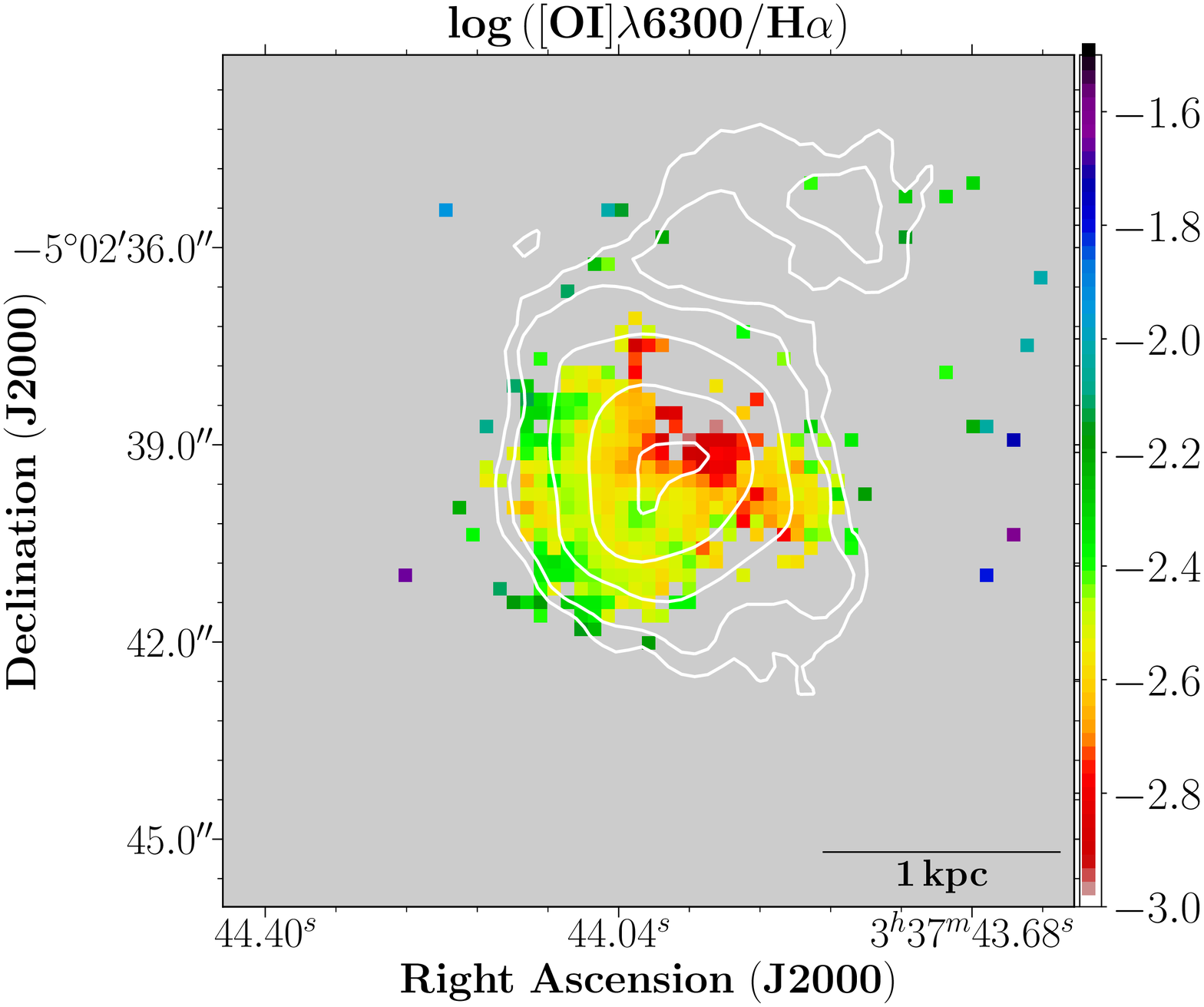}\\
\caption{Line ratio maps of SBS\,0335-052E: [OIII]$\lambda$5007/H$\beta$, [NII]$\lambda$6584/H$\alpha$, [SII]$\lambda$6717,6731/H$\alpha$, [OI]$\lambda$6300/H$\alpha$. Only line fluxes with S/N (per pixel) $>$ 5 are plotted. All maps are presented in logarithmic scale. Isocontours of the HeII$\lambda$4686 emission line flux are shown overplotted for reference. East is left and North is up.}
\label{line_ratio_maps} 
\end{figure*}

The contributions of different ionization mechanisms to the line
emission across a galaxy can be probed quantitatively using diagnostic
line ratios \citep[e.g.,][]{bal81,VO87,kew15}. The three panels of Fig.~\ref{bpt} show
the spatially resolved [N{\sc ii}]$\lambda$6584/H$\alpha$, [S{\sc
ii}]$\lambda\lambda$6717,6731/H$\alpha$, and [O{\sc
i}]$\lambda$6300/H$\alpha$ versus [O{\sc iii}]$\lambda$5007/H$\beta$
diagrams for SBS\,0335-052E, while Fig.~\ref{line_ratio_maps} displays the maps for 
such line ratios. These line ratios are not corrected for reddening, but their reddening
dependence is negligible since they are calculated from
lines which are close in wavelength. These optical diagnostic diagrams are
commonly used to separate gas excitation dominated by massive star
photoionization from other ionizing sources, including AGNs,
post-AGB stars and shocks \citep[e.g.,][]{kew01,kew06,K12,G16,dav17}. In
Fig.~\ref{bpt} each circle represents the line ratios calculated from
the emission line fluxes of an individual spaxel, where the red
circles indicate the HeII-emitting spaxels. The line ratios obtained
from the one-dimensional (1D) spectra of selected regions across our
FOV (see Section~\ref{int} and Table~\ref{table_regions}) are
overplotted on the diagrams as triangles.  The solid black curves
on the three plots define the \cite{kew01}  theoretical upper bound to pure star
formation. The dashed black curve on the [N{\sc
ii}]$\lambda$6584/H$\alpha$-diagram traces the \cite{K03} empirical
classification line, which marks out the upper boundary of the Sloan
Digital Sky Survey star formation sequence.  All spectra lying below
these curves are dominated by star formation.  As indicated in
Fig.~\ref{bpt}, for all positions in SBS\,0335-052E our emission-line
ratios fall in the general locus of star-forming objects according to the
spectral classification scheme proposed by \cite{kew01}
and \cite{K03}. Accordingly, photoionization from hot
massive stars seems to be the dominant excitation mechanism within SBS\,0335-052E,
regardless of the locus in the galaxy. 

Of course, in starburst systems, supernovae remnants (SNRs) and
massive star winds might be present and produce shock-heated gas. The
extension and complex morphology of the ionized gas in SBS\,0335-052E
indeed suggest that the gas excitation and hard
ionization may be partially due to shocks. However, besides the BPT diagram results, our spatially
resolved observations give other reasons why shocks are unlikely to
be the main source of He$^{+}$-ionizing photons in SBS\,0335-052E. We find no sign
of significant [SII]/H$\alpha$ and/or [OI]$\lambda$6300/H$\alpha$
enhancement \citep[a frequent indication of shock-excited gas; e.g.,][]{S85,d96,a08} associated with the HeII-emitting
region (see Fig.\ref{line_ratio_maps}). In particular, there is no [OI]$\lambda$6300
emission with S/N $\geq$ 5 at the location of the HeII shell whose filamentary structure
could, in principle, suggest an important shock excitation (see Fig.\ref{line_maps}); the HeII
shell also shows low values of [SII]/H$\alpha$ associated with (see Fig.\ref{line_ratio_maps}). Our
measured values of [SII]$\lambda\lambda$6717,6731/H$\alpha$ ratio
($\lesssim$ 0.03) in the HeII-emitting spaxels (see Fig.\ref{bpt}) are lower
than the ones observed in SNRs ([SII]$\lambda\lambda$6717,6731/H$\alpha$
$\sim$ 0.5 - 1.0; e.g., \citealt{smith93}) which points against HeII shock
ionization as well. Moreover, the characteristic [OIII] temperature
($\sim$ 20,000 K; e.g., \citealt{I06,P06}) in SBS\,0335-052E is not as high as seen in classical SNRs
\citep[e.g.,][]{b81}.  These observational facts reinforce that notion that the formation
of nebular HeII in SBS\,0335-052E is mainly due to stellar sources, as
mentioned previously. Next, we explore the hot massive
star scenario.

\subsection{Hot stars ionizing continua}

\subsubsection{Models on a star-by-star basis}\label{star_by_star}

For many years, hot Wolf-Rayet stars (WRs) have been proposed to be a
reasonable explanation for HeII excitation, given the good correlation 
between the presence of narrow HeII$\lambda$4686 emission and WR features in
SF systems mostly with metallicities 12+log(O/H)
$\gtrsim$ 8 \citep[e.g.,][]{s96,sv98,b08}. However, this correlation
does not hold in all cases \citep[e.g.,][]{hc07,K11,nm11}. Various studies have shown that
nebular HeII$\lambda$4686 emission does not appear to be always
associated with WRs, especially in metal-deficient objects
\citep[e.g.,][]{g00,TI05,sb12}. In principle, one could envisage that
the lack of WR features in the spectrum of metal-poor galaxies with
nebular HeII$\lambda$4686 emission is likely because of the reduced WR
line luminosities expected at low metallicities
\citep[e.g.,][]{ch06}. Of course, this interpretation might be valid
in some cases, but not at all times; there are examples showing that
the non-detection of WR signatures in some metal-deficient
HeII$\lambda$4686-emitting regions is not an effect of the weakness of
WR bumps, confirming so the lack of connection of HeII$\lambda$4686
emission with the hard radiation from WRs \citep[see][]{K08,K13,sb12,K15}. Thus, the role played by
WRs in ionizing He$^{+}$ at low metallicities is still uncertain. 

\begin{figure*}
\center
\includegraphics[width=8.5cm,clip]{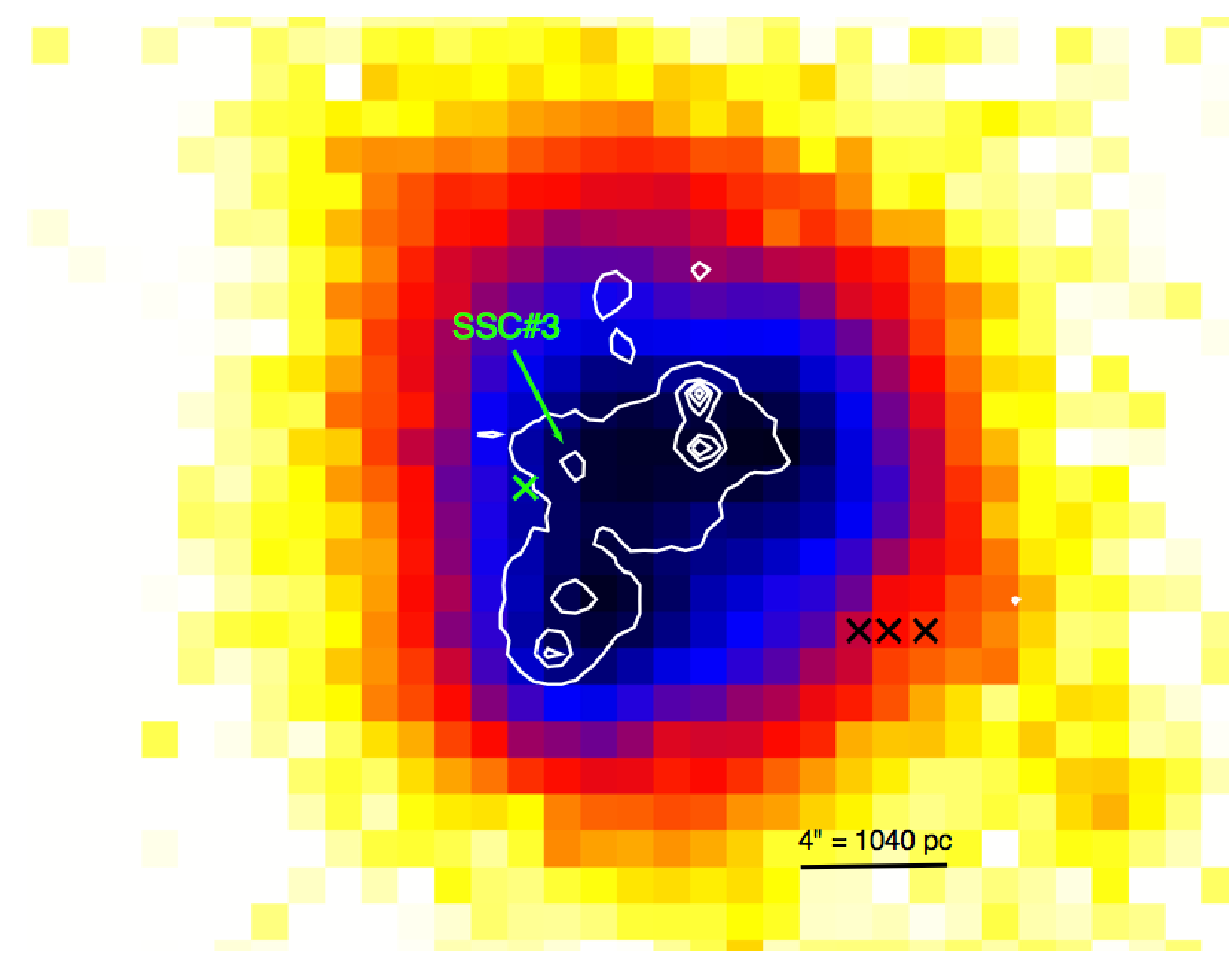}\\
\includegraphics[bb= 18 430 584 714,width=8.5cm,clip]{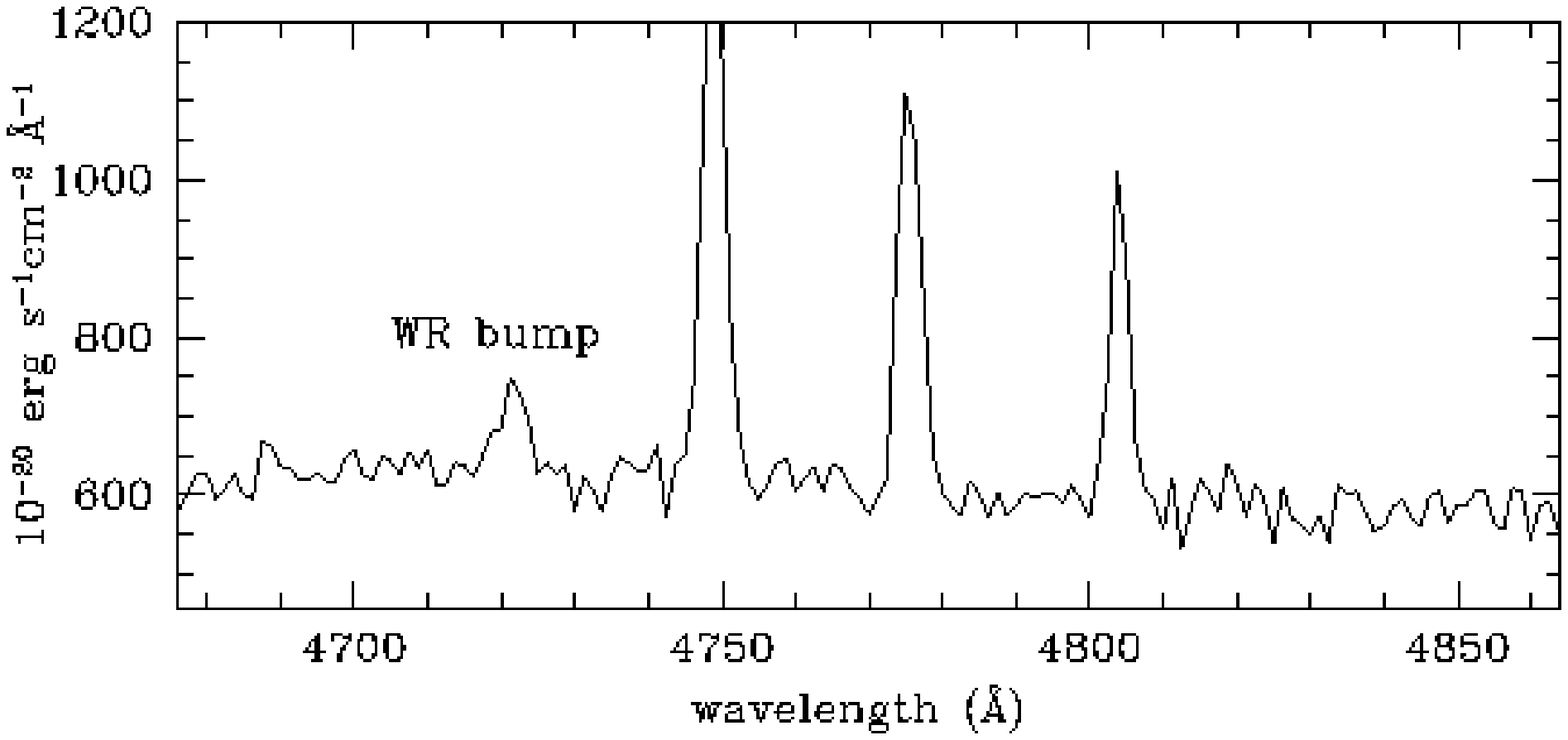}
\includegraphics[bb= 18 430 584 714,width=8.5cm,clip]{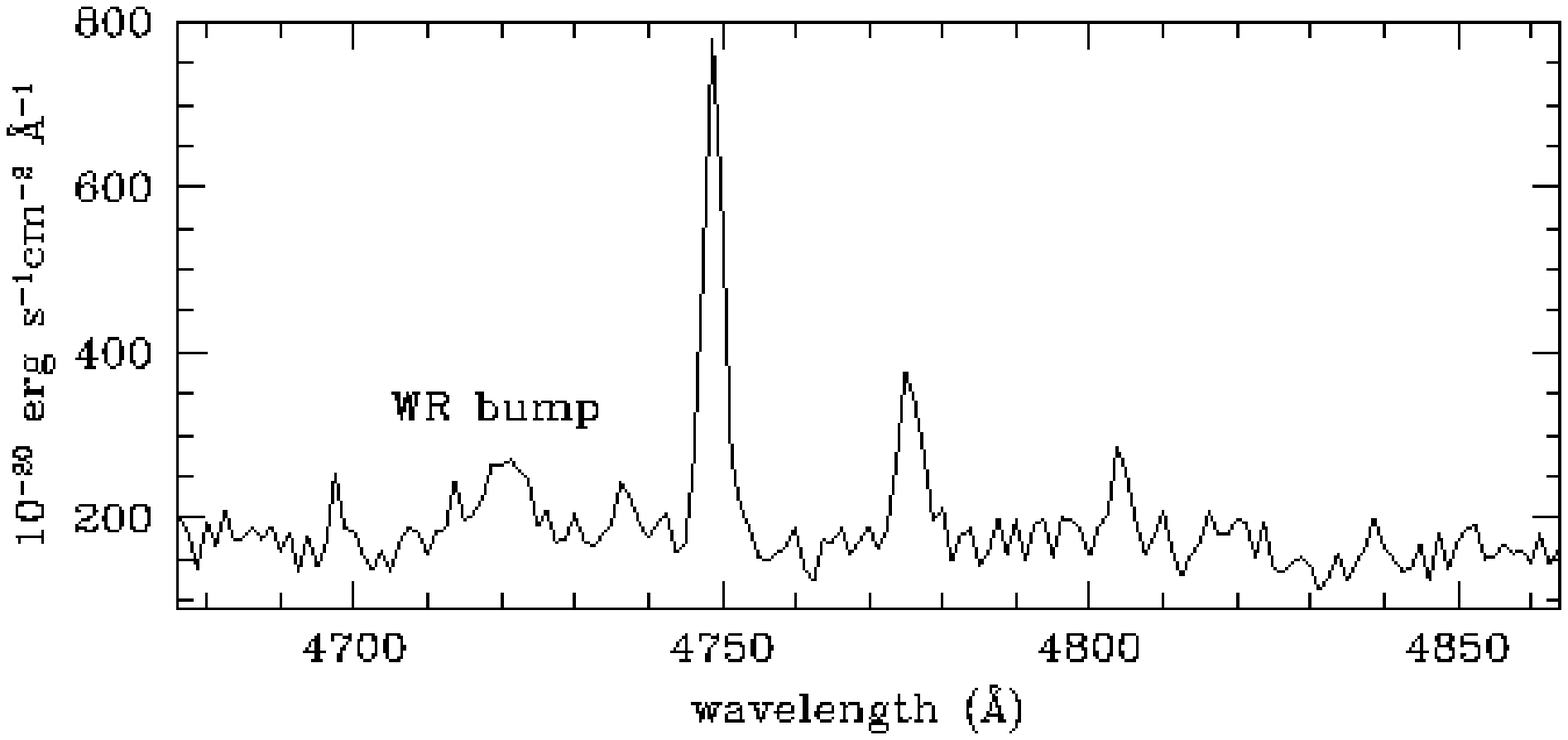}
\caption{{\it Top panel}: Nebular HeII$\lambda$4686 map centered on the
  massive young SSCs whose positions are represented by the white
  countours. The crosses indicate the location of the WR knots
  detected in this work.
  {\it Bottom row}: spectra showing the WR bumps. {\it Bottom-left
    panel}: spectrum of the spaxel nearby SSC\#3, indicated by the green cross in the
  {\it top panel}; {\it Bottom-right panel}: spectrum obtained by adding the emission
from the 3 spaxels marked with black crosses in the {\it top panel}.}
\label{wr} 
\end{figure*}

The existence of a WR stellar population in SBS\,0335-052E, despite
its low Z, was first reported by \cite{I99}. Later, \cite{P06} and
\cite{I06} showed the detection of WRs associated with
SSC \#3.  We searched for WR features in
all 0.2'' $\times$ 0.2'' HeII-emitting spaxels from MUSE.  Our data
set show indications of a WR bump centered at $\lambda$ $\sim$
4656 \AA~ in one spaxel nearby SSC \#3, in agreement with
\cite{P06} and \cite{I06}. In comparison with these works, the
spatially resolved spectroscopy with high spatial resolution of MUSE
allows us to give a more precise location of such WR feature which is
found to be $\sim$ 0.3 arcsec ($\sim$ 78 pc at the distance of
SBS\,0335-052E) southeast of SSC \#3. Moreover, by adding the
emission from three spaxels, we find a new WR detection, also at
$\lambda$ $\sim$ 4656 $\AA$. This new WR knot is located at $\sim$ 1.4
arsec ($\sim$ 360 pc at the distance of SBS\,0335-052E) toward the
west of the center of the SSCs complex\footnote{According to
  \cite{e11}, for type Ib/c SN progenitors, 10 to 20$\%$ of WRs could travel hundreds of pc
  from their birth places. This could be the case of the new WR knot
  detected here, but this discussion is beyond the framework of this paper.}
(see Fig.\ref{wr}). 

The $\lambda$4656 broad bump observed in the spectra of the two WR
knots (see Fig. \ref{wr}) can be read as a signature of early carbon
WRs \citep[WCE; e.g.,][]{ch06,P06}. In IZw18, a nearby galaxy
with extremely low metallicity ($\sim$ 1/40 solar
metallicity; e.g., \citealt{jvm98,K16}), isolated clusters with
WC stars have been detected as well \citep[e.g.,][]{b02}.
Detecting WRs in very metal-deficient
objects is always an important observational constraint because it challenges current stellar evolutionary
models for massive stars, which predict very few, if any, 
WRs in extremely metal-poor environments \citep[e.g.,][]{L14,bpass}.
Here, the total flux of the WCE bump measured 
from each WR knot spectrum provides a similar reddening-corrected
luminosity of L$_{WCE}$ $\sim$ 2.4 $\times$
10$^{36}$ erg s$^{-1}$. 
\cite{ch06} computed WCE (WC4) models at the
LMC and IZw18 metallicities, i.e., 12+log(O/H) $\sim$ 8.36 \citep[e.g.,][]{rd90} and 7.11 \citep[e.g.,][]{K16},
respectively. 

As the metallicity of SBS\,0335-052E [12+log(O/H)
$\sim$ 7.3]  is closer to that of 
IZw18, we compared our observations with the IZw18-like WCE models
to derive the number of WRs in SBS\,0335-052E. Taking the
luminosity of 3.5 $\times$ 10$^{35}$ erg s$^{-1}$ for a single
IZw18-like WC4 star \citep[][]{ch06}, this corresponds to an equivalent number of
$\sim$ 7 WCEs in each WR knot, i.e., a total of 14 WCEs in
SBS\,0335-052E. From these 14 WCEs, a total flux of Q(HeII) = 4.4 $\times$ 10$^{48}$
photon s$^{-1}$ is expected (assuming Q(HeII) = 10$^{47.5}$ photon
s$^{-1}$ for one IZw18-like WCE; \citealt{ch06}). This is about 500 times
lower than the Q(HeII)$_{HeII-MB}$ = 2.25 $\times$ 10$^{51}$ 
for the HeII main body region, and a factor of $\sim$ 700 lower than
the Q(HeII)$_{Int}$ = 3.17 $\times$ 10$^{51}$ derived from our data.

Based on the WCE models by \cite{ch06}, therefore, more than 7000 WRs are
required to explain the HeII-ionization budget measured. Such a very
large WR population is not supported by state-of-the-art stellar
evolutionary models for (single and binary)
massive stars in metal-poor environments \citep[see e.g.,][and
references therein]{L14,bpass}, and it is not
compatible with the total stellar mass of the six SSCs \citep[M$_{\star,SSCs}$ $\sim$ 5.6
$\times$ 10$^{6}$ M$_{\odot}$;][]{R08} where the star formation in SBS\,0335-052E is
mostly concentrated; assuming a Salpeter initial mass
function \citep[IMF;][]{salpeter} with M$_{up}$ = 150 M$_{\odot}$ (M$_{up}$ = 300
M$_{\odot}$), a cluster with $>$ 7 (5) times the SSCs total mass would
be needed to produce $>$ 7000 WRs in SBS\,0335-052E. 
Moreover, the nebular HeII$\lambda$4686 emission is by far more
extended than the places occupied by the WR knots (see Fig.\ref{wr}).   
Based on IFS studies, we have observed similar cases of spatial offset between WRs
and HeII$\lambda$4686-emitting regions for other nearby, metal-poor SF
galaxies \citep[see][]{K08,K13,K15}.  All this indicates that,
although the contribution from WRs to HeII-ionization cannot be
disregarded, they are not exclusively responsible for the
HeII$\lambda$4686 emission in SBS\,0335-052E, in line with other
metal-deficient galaxies as mentioned above. The 
evolution of massive stars into WR stars is not well understood, and the number of WRs derived
here is limited to the current available models for metal-poor WRs
which suffer from still unresolved uncertainties \citep[see e.g.,][]{smith02,vk05,ch06,PC07,mm12}.  

Considering hot massive stars to be the primary source of
He$^{+}$-ionizing photons in SBS\,0335-052E, as our IFS observations
indicate, we need to investigate other types of stars besides
WRs. 
Although classical nebulae ionized by O stars having Teff $<$ 55,000 K are
not expected to produce strong He$^{++}$, certain O stars are predicted to be hot enough to ionize
HeII \citep[][]{k02}. According to the hottest models (Te = 60,000K) of massive O stars at
Z $\sim$ 3$\%$ (i.e, the SBS\,0335-052E metallicity) by \citet{k02},
we infer that $\sim$ 16,000 stars with 150 
M$_{\odot}$ (with predicted Q(HeII) $\approx$ 1.9 $\times$ 10$^{47}$
photon s$^{-1}$ each; see Fig.12 from \citealt{k02})
are required to explain the derived Q(HeII)$_{Int}$
budget. Alternatively, if we take the 300 M$_{\odot}$ star hottest
models (with Q(HeII) $\approx$ 9 $\times$ 10$^{48}$
photon s$^{-1}$ each), $\sim$ 360 of these stars would be needed to
produce the observed Q(HeII)$_{Int}$. For a Salpeter IMF, 16,000 stars
with 145 $\leq$ M$_{\star}$/M$_{\odot}$ $\leq$ 155 would imply a cluster with
$>$ 270 $\times$ M$_{\star,SSCs}$; on the other hand, 360 stars with
290 $\leq$ M$_{\star}$/M$_{\odot}$ $\leq$ 310 requires a cluster mass of
$\sim$ 16 $\times$ M$_{\star,SSCs}$. A Kroupa
IMF \citep[][]{kroupa} does not solve the cluster mass issue neither.  

Work in recent years has shown that rotation is a key ingredient in shaping the
evolution of massive stars with very low metallicities \citep[Z $<$
SMC metallicity down to Population III stars; see][and references
therein]{mm17}. Fast rotators (with initial {\it v$_{rot}$} $\gtrsim$ 300 km
s$^{-1}$) are expected to lead to chemically homogeneous
evolution (CHE) in which the star becomes brighter and hotter, and
thus more ionizing photons, specially in the extreme UV, are emitted than in the corresponding
non-rotating case
\citep[e.g.,][]{brott11,levesque12,yoon}. Presently, although we still
know little about rotation velocities of massive stars and their
variation with environment, observations seem to favor fast rotators
at low Z \citep[e.g.,][]{m07,hunter}. Model predictions suggest that
the effects of rotation, like CHE, should be enhanced at lower
metallicities.
There is an increase of theoretical and observational evidence
which supports the significant role of rotation among the generations of first stars,
with Z=0 or extremely low metallicities, and fast rotating massive stars were likely common phenomena in the early
universe \citep[e..g,][]{l08,c08,mm12,yoon}. Modern models for
low-metallicity massive fast-rotating single stars which undergo CHE
have been published by \citet{dori};
the authors called them transparent wind ultraviolet intense stars (or
TWUIN stars). Considering
all this and the extremely low-Z of SBS\,0335-052E, we compare our observations
with the TWUINs predictions\footnote{We note that the TWUIN models by
\citet{dori} are predicted at the IZw18 metallicity, i.e., slightly lower than
the SBS\,0335-052E metallicity.}, following the same approach that we used
previously for non-rotating WCE and massive O stellar models. Taking
the computed Q(HeII) = 7.37 $\times$ 10$^{48}$ photon s$^{-1}$  for
the most massive 294 M$_{\odot}$ TWUIN, we require that $\sim$ 430
($\sim$ 300) such stars are necessary to explain the Q(HeII)$_{Int}$
(Q(HeII)$_{HeII-MB}$). Again, hundreds of these super-massive TWUINs do not
match the M$_{\star,SSCs}$ of SBS\,0335-052E. If we instead, apply the
same approach using lower mass TWUIN models, it makes even harder to account for the observations. Besides the Q(HeII) budget, we have measured values of 
HeII$\lambda$4686/H$\beta$ as high as 0.06 within the HeII$\lambda$4686 main body
region. Under ionization-bounded conditions, even
the most massive TWUIN models cannot reproduce these values of
HeII$\lambda$4686/H$\beta$ (see Table B.1 from \citet{dori}). 

As mentioned in Section~\ref{intro}, searches for very metal-poor
starbursts and PopIII-hosting galaxies at high-z using HeII lines have
been performed, and is among the main {\it JWST} science drivers
\citep[e.g.,][]{sch08,c13,vhb2015}. This search is based on the
predicted high effective temperature for PopIII stars which will emit
a large number of photons with energy $>$ 54eV, and also on the
expected and observed increase of the nebular HeII line with
decreasing Z \citep[e.g.,][]{g00,sch03,sch08,b08}. So a speculative
possibility to explain the derived Q(HeII) in SBS\,0335-052E could be
based on nearly metal-free ionizing stars.  We have used recent models
for fast-rotating Z=0 stars \citep[][]{yoon} as an approximation for
the HeII ionizing output of nearly metal-free stars. According to
these Z=0 models, we found that a significantly smaller number of such
stars, in comparison with the models considered previously, could
explain the observed HeII budget; $\sim$ 160 (230) stars with mass
M$_{ini}$ = 150 M$_{\odot}$ (with Q(HeII) $\approx$ 1.42 $\times$
10$^{49}$ photon s$^{-1}$ each) would be sufficient to account for the
Q(HeII)$_{HeII-MB}$ (Q(HeII)$_{Int}$). Additionally, the ionizing
spectra produced by these star models are hard enough to explain the
highest He II$\lambda$4686/H$\beta$ values observed, assuming that
ionization-bounded conditions are met. Further speculation on the Z=0
stars hypothesis is beyond the scope of this paper. 

Despite its nature as an entirely hypothetical scenario, it might suggest that
stars with metallicity considerably lower than that of the HII regions
in SBS\,0335-052E are required to produce the total HeII ionizing
photon flux (see next Section).

\subsubsection{Stellar Population Synthesis Models}\label{binary}

The effects of stellar multiplicity have been
recognized to play a key role when modelling young stellar populations
\citep[see][and references therein]{bpass}. Recently, it has been
suggested that most massive stars evolve as part of binary systems
\citep[e.g.,][]{langer12,sana12,sana14}.  At low metallicities,
different stellar evolution pathways, including those resulting from
interacting binaries, are expected to become more and more important \citep[e.g.,][]{z13,s14,s16}. According to
population synthesis codes incorporating binary interactions, binarity
leads to a bluer stellar population, and extends the lifetime of the WR
stars up to ages later than 10 Myrs. This allows harder ionizing
photons to exist at later times than expected from a single star
population \citep[e.g.,][]{belkus,z15,bpass}.

Here we considered a suite of models from the most recent data release
of {\it Binary Population and Spectral Synthesis}
\citep[BPASSv2.1;][]{bpass}. By integrating the predicted BPASS SEDs
over the wavelength range 1 \AA
$<$ $\lambda$ $<$ 228 \AA, we computed Q(HeII) for all IMFs available on the BPASSv2.1 data release. For each IMF we took into account the three lowest metallicity models
: Z=10$^{-3}$, 10$^{-4}$, 10$^{-5}$. The models at Z=10$^{-3}$
correspond to the metallicity closest to that of SBS\,0335-052E. A
fundamental benefit of the 
new BPASSv2.1 is that now they have spectra at the lowest
metallicities Z = 10$^{-5}$ and 10$^{-4}$. We took advantage of this
since the influence of 
massive stars on the synthetic spectrum is believed to be
strongest at these low metallicities. Finally, for every set of models, we also
considered the BPASS predictions which do not include binary
evolution, i.e., the BPASS single-star models. Thus this set of models
allows us to check the impact on the HeII ionizing flux
output by varying the IMF, metallicity and binarity. Our estimates assume an instantaneous burst at
10$^{6.8}$ yr because that model age matches the
SSCs mean age of $\sim$ 6 Myr \citep[][]{T97,R08}. The results obtained from the BPASSv2.1
models considered here are summarized in Table~\ref{tab_bpass}. 

From Table~\ref{tab_bpass}, we see that, at a given IMF, the predicted
Q(HeII) always increases with decreasing metallicity, both for
single-star and binary models. This effect is expected since stars at
lower metallicity have hotter effective temperatures which boost the
ionizing photon flux. We also found that the Q(HeII) output of a
binary population exceeds that of the single star population at all
metallicities, regardless the IMF adopted. This outcome agrees
with \cite{bpass} which predict a longer ionizing photon production
lifetime of the binary stars overtaking the single stars at late ages
for any ongoing or evolving starburst. Scaling the Q(HeII) values from
Table~\ref{tab_bpass} to the M$_{\star,SSCs}$ of $\sim$ 5.6 $\times$
10$^{6}$ M$_{\odot}$ we infer that none of the models with Z=10$^{-3}$
(the Z that better match the SBS\,0335-052E metallicity of $\sim$ 3-4
$\%$ solar) can reproduce the observed integrated Q(HeII)$_{Int}$ =
3.17 $\times$ 10$^{51}$ photon s$^{-1}$.  At Z=10$^{-3}$, predicted
Q(HeII) reaches its maximum for the binary star model with IMF
``imf100-300''; this value is $\sim$ 1.3 $\times$ 10$^{50}$ photon
s$^{-1}$ (= 2.39 $\times$ 10$^{49}$ photon s$^{-1}$ scaled to the
M$_{\star,SSCs}$; see Table~\ref{tab_bpass}), i.e., $\sim$ 24 times
lower than the derived Q(HeII)$_{Int}$.

The only two models that can explain the derived Q(HeII)$_{Int}$
budget are the binary ones at Z=10$^{-5}$ for a shallower 'top-heavy'
IMF with either M$_{up}$=100 M$_{\odot}$ (``imf100-100'') or
M$_{up}$=300 M$_{\odot}$ (``imf100-300''); both predict a similar Q(HeII) of $\sim$ 4.4
$\times$ 10$^{51}$ photon s$^{-1}$ or 4.1 $\times$ 10$^{51}$ photon
s$^{-1}$ (after scaling to the M$_{\star,SSCs}$;
see Table~\ref{tab_bpass}), respectively. These predictions 
exceed the observed Q(HeII)$_{Int}$ which could suggest that some
fraction of the He$^{+}$-ionizing photons may be able to escape the
galaxy. Previous work claim that up to $\sim$ 50 $\%$ of the hydrogen
ionizing flux could be leaking out the HII regions of SBS\,0335-052E,
possibly due to clumping and filaments in its ISM \citep[see
e.g.,][]{R08,J09,H17}. These structures in the ionized gas might be acting as
escape pipes for the HeII Lyman continuum as well.

Thus, the outcomes from the BPASSv2.1 code indicate that while calling upon
super-massive stars (M $>$100 M$_{\odot}$) is not essential, a binary
population distributed over a 'top-heavy' IMF seem to be required to clear up the observed HeII flux
budget. More interestingly, according to these modern models,
stars with metallicity (Z = 0.05 $\%$ solar) much lower than that of
the ionized gas (Z $\sim$ 3-4 $\%$ solar) of SBS\,0335-052E are to be
invoked to explain the HeII ionization in this galaxy. We highlight
that similar results have been found to account for by the HeII
ionization in other nearby, metal-poor SF galaxies
\citep[see][]{K15,S17}.  Certainly, in this
hypothetical scenario, such very metal-poor stars cannot belong to the
same SSCs which host more chemically evolved objects. Further investigation is needed to understand
the origin of this apparent inconsistency. Nevertheless, we should
bear in mind the uncertainties yet unsolved in the BPASS code; the
reader is referred to Section 7 from \cite{bpass} for details. Also,
caution should be exercised when evaluating the Z = 10$^{-5}$ and
10$^{-4}$ models; empirical data for stellar atmospheres at these low
Z are not available, and thus consist of theoretical extrapolation
from higher-Z atmospheres which are better tested.

Although a detailed comparison between atmosphere and wind models is beyond the
scope of this work, intermediate-mass stripped stars 
could be an important source of hard ionizing radiation.  
Such stars, thought to be very hot objects emitting the
majority of their photons in the extreme UV, have been stripped of
their envelope through interaction with a binary companion 
\citep{g17}.
\cite{g18} computed atmosphere models for these stars, but
they are not yet considered in BPASS, and thus we have not considered them here.
Although models suggest that their importance decreases with decreasing
metallicity and thus should not be relevant for extremely metal-poor objects, 
the Q(HeII) photon flux predictions of such stars
are still uncertain \citep{g18}, and any conclusion about
their role in galaxies like SBS0335-052 would be premature. 

\begin{table*}
\caption{Integrated Q(HeII) from BPASSv2.1 for instantaneous starburt models of initial
  stellar mass 10$^{6}$ M$_{\odot}$ at 10$^{6.8}$ yr}
\label{tab_bpass}
\centering
\begin{minipage}{14.9cm}
\centering
\begin{tabular}{|l|l|cc|}
\hline\hline
IMF \footnote{IMFs are labelled following the BPASSv2.1 designation \citep[][]{bpass}}  & Z (Z/Z$_{\odot}$) \footnote{BPASS models assume Z$_{\odot}$=0.02 for consistency with their empirical mass-loss rates which were
scaled from this value.} & \multicolumn{2}{c|}{Q(HeII) photon s$^{-1}$}  
\\ & & Single & Binary    
\\ \hline
\multirow{3}{*}{imf100-100\footnote{imf100-100: $\alpha_{1}$=-1.30 (0.1-0.5 M$_{\odot}$), $\alpha_{2}$=-2.00
    (0.5-100 M$_{\odot}$); imf100-300: as imf100-100 but with
    M$_{up}$=300 M$_{\odot}$}}  &  10$^{-5}$ (0.05 $\%$)   &1.40$\times$ 10$^{49}$  &7.89$\times$ 10$^{50}$  \\
                                    & 10$^{-4}$ (0.5 $\%$)   &6.45$\times$ 10$^{48}$  & 3.40$\times$ 10$^{50}$ \\ 
                                    & 0.001 (5 $\%$)  & 5.05$\times$ 10$^{46}$ & 2.18$\times$ 10$^{49}$\\ \hline

\multirow{3}{*}{imf100-300$^{c}$}   & 10$^{-5}$ (0.05 $\%$) &1.20$\times$ 10$^{49}$  &7.37$\times$ 10$^{50}$   \\
                                   & 10$^{-4}$ (0.5 $\%$) &6.16$\times$ 10$^{48}$  &3.48$\times$ 10$^{50}$  \\
                                   & 0.001 (5 $\%$) &4.30$\times$ 10$^{46}$ &2.39$\times$ 10$^{49}$\\ \hline

\multirow{3}{*}{imf135all-100\footnote{Salpeter IMF with M$_{up}$=100 M$_{\odot}$}}   & 10$^{-5}$ (0.05 $\%$)   & 5.01$\times$ 10$^{48}$  & 2.09$\times$ 10$^{50}$   \\
                                                       & 10$^{-4}$ (0.5 $\%$)  & 2.29$\times$ 10$^{48}$  & 8.97$\times$ 10$^{49}$  \\
                                                       &  0.001 (5 $\%$)   &  1.77$\times$ 10$^{46}$  &  5.94$\times$ 10$^{48}$ \\  \hline 
\multirow{3}{*}{imf135-100\footnote{These are the default IMFs
    recommended by \cite{bpass}; imf135-100: $\alpha_{1}$=-1.30 (0.1-0.5 M$_{\odot}$), $\alpha_{2}$=-2.35
    (0.5-100 M$_{\odot}$); imf135-300: as imf135-100 but with
    M$_{up}$=300 M$_{\odot}$}}    &  10$^{-5}$ (0.05 $\%$)  &6.87$\times$ 10$^{48}$  &2.71 $\times$ 10$^{50}$  \\
                                    & 10$^{-4}$ (0.5 $\%$)        &3.10$\times$ 10$^{48}$  & 1.16$\times$ 10$^{50}$ \\ 
                                    & 0.001 (5 $\%$)        & 2.40$\times$ 10$^{46}$ & 7.68$\times$ 10$^{48}$\\ \hline

\multirow{3}{*}{imf135-300$^{e}$}   & 10$^{-5}$ (0.05 $\%$) &6.59$\times$ 10$^{48}$  &2.79$\times$ 10$^{50}$   \\
                                   & 10$^{-4}$ (0.5 $\%$) &2.98$\times$ 10$^{48}$  &1.27$\times$ 10$^{50}$  \\
                                   & 0.001 (5 $\%$) &2.30$\times$ 10$^{46}$ &8.99$\times$ 10$^{48}$\\ \hline
\multirow{3}{*}{imf170-100\footnote{imf170-100: $\alpha_{1}$=-1.30 (0.1-0.5 M$_{\odot}$), $\alpha_{2}$=-2.70
    (0.5-100 M$_{\odot}$); imf170-300: as imf170-100 but
    withM$_{up}$=300 M$_{\odot}$}}    
                                     &  10$^{-5}$ (0.05 $\%$)   &2.60$\times$ 10$^{48}$  & 7.22$\times$ 10$^{49}$  \\
                                    & 10$^{-4}$ (0.5 $\%$)        &1.16$\times$ 10$^{48}$  & 3.07$\times$ 10$^{49}$ \\ 
                                    & 0.001 (5 $\%$)        & 8.77$\times$ 10$^{45}$ & 2.13$\times$ 10$^{48}$\\ \hline
\multirow{3}{*}{imf170-300$^{f}$}    
                                     &  10$^{-5}$ (0.05 $\%$)   &2.58$\times$ 10$^{48}$  & 7.62$\times$ 10$^{49}$  \\
                                    & 10$^{-4}$ (0.5 $\%$)       &1.15$\times$ 10$^{48}$  & 3.43$\times$ 10$^{49}$ \\ 
                                    & 0.001 (5 $\%$)        & 8.70$\times$ 10$^{45}$ & 2.46$\times$ 10$^{48}$\\ \hline
\hline
\end{tabular}
\end{minipage}
\end{table*}

\section{Summary and concluding remarks}

Observational data of high-ionization lines, like HeII, in the
reionization era have been recently accumulated, and there is a growing body of
evidence that HeII-emitters are more frequent among high-z
galaxies. Narrow HeII emission has been claimed to be a good tracer of the
elusive PopIII-stars. The HeII line is in comfortable reach of next generation telescopes,
like {\it JWST} and {\it ELT}, which will detect the rest-frame UV of thousands of
galaxies during the epoch of reionization. In light of these new and upcoming
observations, a more sophisticated understanding of the
high-ionization phenomenon at z $\sim$ 0 is pivotal to interpret the data in a
physically meaningful manner, and to possibly constrain sources
responsible for the H and HeII reionization epochs. Optical MUSE-IFU and X-ray Chandra observations of the nearby,
extremely metal-poor bursty-galaxy SBS\,0335-052E with nebular HeII emission are presented in
this work. Such objects are excellent primordial analogues that can
provide clues to the physical properties of galaxies in the early universe.

Based on the MUSE data we created spectral maps of relevant emission
lines in the optical range. These data provide us with a new 2D view
of the ionized ISM in SBS\,0335-052E, in particular the
high-ionization nebular HeII$\lambda$4686 line \citep[see also][]{H17}. We find a highly
extended HeII$\lambda$4686-emitting region reaching distances
$\gtrsim$ 1.5 Kpc from the youngest SSCs. The HeII$\lambda$4686 map
shows three maxima which are offset from the brightest UV SSCs.  We
took benefit of our IFU data to create 1D spectra of several regions
based on the spatial distribution of the HeII$\lambda$4686
emission. For the first time, using the derived SBS\,0335-052E
integrated spectrum, we recovered the entire nebular HeII emission and
compute the corresponding total HeII-ionization budget,
Q(HeII)$_{int}$ $\approx$ 3.17 $\times$ 10$^{51}$ photon s$^{-1}$. As we showed in
\citet{K15}, this observational quantity is essential to perform a
free-aperture investigation on the formation of narrow HeII line.

From our analysis of the combined MUSE and Chandra data, we infer that
the Q(HeII) provided by the X-ray emission from SBS\,0335-052E is well
below the needed value to ionize He$^{+}$ at the level measured in
this galaxy. Although we cannot discard some contribution to the HeII
excitation from shocks, our spaxel-by-spaxel study from MUSE favors
hot massive stars as the main agent of the HeII ionization in
SBS\,0335-052E. 

Using the MUSE observations we located two WR knots from which one is
identified for the first time here. The WR knot spectra reveal the
presence of carbon-type WR stars in SBS\,0335-052E. In comparison with
long-slit or single-fiber spectroscopy, the IFS technique can enhance the contrast of
the WR bump emission against the galaxy continuum. This minimizes the
WR bump dilution, also allowing the identification of WR stars where they were not
detected before \citep[see also][]{K08,K13}.
To investigate the stellar source scenario we compare
observations with stellar models applying two approaches; a
more simple one which makes use of models of (rotating and non-rotating) single massive stars on a
star-by-star basis, and a second one based on the new release of
stellar population synthesis BPASSv2.1. We demonstrated that both approaches have merits, and taken together,
they can provide guidance to interpret the observations. The star-by-star study points out that WRs solely
cannot explain the HeII-ionizing energy budget derived for
SBS\,0335-052E. From the BPASSv2.1 code we show that only the binary models with a
'top-heavy' IMF (i.e. non-standard IMF) at Z=10$^{-5}$ (i.e., Z $\sim$
70 times lower than the Z of the ionized gas in SBS\,0335-052E)
) provide sufficient He$^{+}$-ionizing photons to explain the observed
Q(HeII)$_{Int}$. The approach on a  star-by-star basis also
suggests that stars more metal-deficient than the galaxy main body are
required. Investigating such metallicity discrepancy at z$\sim$ 0 feels the
necessity for additional research, and it is fundamental to interpret
high-ionization emission in the distant universe.

This paper shows the need to identify nearby, metal-poor
HeII-emitters, like SBS\,0335-052E. IFS studies of such galaxies enable
extended insight into their “realistic” ISM and massive stars,
therefore providing constraints on high-redshift galaxy evolution, and
on metal-poor stellar models. The work presented here can guide through the
preparation of forthcoming searches for primeval objects,
one of the main science drivers for next-generation telescopes.

\section*{Acknowledgements}
We are very grateful to our referee for providing constructive comments
and help in improving the manuscript.
This work has been partially funded by research project
AYA2017-79724-C4-4-P from the Spanish PNAYA.  CK, JVM, SD and JIP
acknowledge financial support from Junta de Andalucia Excellence
Project PEX2011-FQM705. MAG acknowledges support of the grant AYA
2014-57280-P, co-funded with FEDER funds. LKH thanks the IAA-CSIC for warm hospitality in the visit during which the work for this paper was initiated. This work made use of the
v2.1 of the Binary Population and Spectral Synthesis (BPASS) models as
last described in \cite{bpass}.

%

\bibliographystyle{mn2e}

\label{lastpage}
\end{document}